\documentclass[
    journal,
    twocolumn,
    ]{IEEEtran}
\usepackage{graphicx}
\usepackage{amsmath}
\usepackage{amssymb,mathtools}
\usepackage{amsthm}
\usepackage{verbatim}
\usepackage{psfrag}
\usepackage{color}
\usepackage{url}
\usepackage{cite}
\usepackage{epsfig}
\usepackage{epstopdf}
\usepackage{footmisc}
\usepackage{stackrel}
\usepackage{float}
\usepackage{enumerate}
\usepackage{graphicx}
\usepackage{subcaption}
\usepackage{graphicx}

\usepackage[nolist]{acronym}
\begin{document}
\begin{acronym}
	
\acro{AC}{address coding}
\acro{ACF}{autocorrelation function}
\acro{ACR}{autocorrelation receiver}
\acro{ADC}{analog-to-digital converter}
\acrodef{aic}[AIC]{Analog-to-Information Converter}     
\acro{AIC}[AIC]{Akaike information criterion}
\acro{aric}[ARIC]{asymmetric restricted isometry constant}
\acro{arip}[ARIP]{asymmetric restricted isometry property}

\acro{ARQ}{automatic repeat request}
\acro{AUB}{asymptotic union bound}
\acrodef{awgn}[AWGN]{Additive White Gaussian Noise}     
\acro{AWGN}{additive white Gaussian noise}

\acro{APSK}[PSK]{asymmetric PSK} 

\acro{waric}[AWRICs]{asymmetric weak restricted isometry constants}
\acro{warip}[AWRIP]{asymmetric weak restricted isometry property}
\acro{BCH}{Bose, Chaudhuri, and Hocquenghem}        
\acro{BCHC}[BCHSC]{BCH based source coding}
\acro{BEP}{bit error probability}
\acro{BFC}{block fading channel}
\acro{BG}[BG]{Bernoulli-Gaussian}
\acro{BGG}{Bernoulli-Generalized Gaussian}
\acro{BPAM}{binary pulse amplitude modulation}
\acro{BPDN}{Basis Pursuit Denoising}
\acro{BPPM}{binary pulse position modulation}
\acro{BPSK}{binary phase shift keying}
\acro{BPZF}{bandpass zonal filter}
\acro{BSC}{binary symmetric channels}              
\acro{BU}[BU]{Bernoulli-uniform}
\acro{BER}{bit error rate}

\acrodef{cdf}[CDF]{cumulative distribution function}   
\acro{CDF}{cumulative distribution function}
\acrodef{c.d.f.}[CDF]{cumulative distribution function}
\acro{CCDF}{complementary cumulative distribution function}
\acrodef{ccdf}[CCDF]{complementary CDF}               
\acrodef{c.c.d.f.}[CCDF]{complementary cumulative distribution function}
\acro{CD}{cooperative diversity}

\acro{CDMA}{code division multiple access}
\acro{ch.f.}{characteristic function}
\acro{CIR}{channel impulse response}
\acro{cosamp}[CoSaMP]{compressive sampling matching pursuit}
\acro{CR}{cognitive radio}
\acro{cs}[CS]{compressed sensing}                   
\acrodef{cscapital}[CS]{Compressed sensing}
\acrodef{CS}[CS]{compressed sensing}
\acro{CSI}{channel state information}

\acro{CCSDS}{consultative committee for space data systems}
\acro{CC}{convolutional coding}

\acro{DAA}{detect and avoid}
\acro{DAB}{digital audio broadcasting}
\acro{DCT}{discrete cosine transform}
\acro{dft}[DFT]{discrete Fourier transform}
\acro{DR}{distortion-rate}
\acro{DS}{direct sequence}
\acro{DS-SS}{direct-sequence spread-spectrum}
\acro{DTR}{differential transmitted-reference}
\acro{DVB-H}{digital video broadcasting\,--\,handheld}
\acro{DVB-T}{digital video broadcasting\,--\,terrestrial}

\acro{ECC}{European Community Commission}
\acro{EED}[EED]{exact eigenvalues distribution}
\acro{ELP}{equivalent low-pass}

\acro{FC}[FC]{fusion center}
\acro{FCC}{Federal Communications Commission}
\acro{FEC}{forward error correction}
\acro{FFT}{fast Fourier transform}
\acro{FH}{frequency-hopping}
\acro{FH-SS}{frequency-hopping spread-spectrum}
\acrodef{FS}{Frame synchronization}
\acro{FSsmall}[FS]{frame synchronization}

\acro{GA}{Gaussian approximation}
\acro{GF}{Galois field }
\acro{GG}{Generalized-Gaussian}
\acro{GIC}[GIC]{generalized information criterion}
\acro{GLRT}{generalized likelihood ratio test}
\acro{GPS}{Global Positioning System}

\acro{HAP}{high altitude platform}

\acro{IDR}{information distortion-rate}
\acro{IFFT}{inverse fast Fourier transform}
\acro{iht}[IHT]{iterative hard thresholding}
\acro{i.i.d.}{independent, identically distributed}
\acro{IoT}{Internet of Things}                      
\acro{IR}{impulse radio}
\acro{lric}[LRIC]{lower restricted isometry constant}
\acro{lrict}[LRICt]{lower restricted isometry constant threshold}
\acro{ISI}{intersymbol interference}

\acro{LEO}{low earth orbit}
\acro{LF}{likelihood function}
\acro{LLF}{log-likelihood function}
\acro{LLR}{log-likelihood ratio}
\acro{LLRT}{log-likelihood ratio test}
\acro{LOS}{line-of-sight}
\acro{LRT}{likelihood ratio test}
\acro{wlric}[LWRIC]{lower weak restricted isometry constant}
\acro{wlrict}[LWRICt]{LWRIC threshold}
\acro{LPWAN}{low power wide area network}
\acro{LoRaWAN}{Low power long Range Wide Area Network}

\acro{MB}{multiband}
\acro{MC}{multicarrier}
\acro{MDS}{mixed distributed source}
\acro{MF}{matched filter}
\acro{m.g.f.}{moment generating function}
\acro{MI}{mutual information}
\acro{MIMO}{multiple-input multiple-output}
\acro{MISO}{multiple-input single-output}
\acrodef{maxs}[MJSO]{maximum joint support cardinality}                       
\acro{ML}[ML]{maximum likelihood}
\acro{MMSE}{minimum mean-square error}
\acro{MMV}{multiple measurement vectors}
\acrodef{MOS}{model order selection}
\acro{M-PSK}[${M}$-PSK]{$M$-ary phase shift keying}                       
\acro{M-APSK}[${M}$-PSK]{$M$-ary asymmetric PSK} 

\acro{M-QAM}[$M$-QAM]{$M$-ary quadrature amplitude modulation}
\acro{MRC}{maximal ratio combiner}                  
\acro{maxs}[MSO]{maximum sparsity order}                                      
\acro{M2M}{machine to machine}                                                
\acro{MUI}{multi-user interference}

\acro{NB}{narrowband}
\acro{NBI}{narrowband interference}
\acro{NLA}{nonlinear sparse approximation}
\acro{NLOS}{non-line-of-sight}
\acro{NTIA}{National Telecommunications and Information Administration}

\acro{OC}{optimum combining}                             
\acro{OC}{optimum combining}
\acro{ODE}{operational distortion-energy}
\acro{ODR}{operational distortion-rate}
\acro{OFDM}{orthogonal frequency-division multiplexing}
\acro{omp}[OMP]{orthogonal matching pursuit}
\acro{OSMP}[OSMP]{orthogonal subspace matching pursuit}
\acro{OQAM}{offset quadrature amplitude modulation}
\acro{OQPSK}{offset QPSK}

\acro{OQPSK/PM}{OQPSK with phase modulation}

\acro{PAM}{pulse amplitude modulation}
\acro{PAR}{peak-to-average ratio}
\acrodef{pdf}[PDF]{probability density function}                      
\acro{PDF}{probability density function}
\acrodef{p.d.f.}[PDF]{probability distribution function}
\acro{PDP}{power dispersion profile}
\acro{PMF}{probability mass function}                             
\acrodef{p.m.f.}[PMF]{probability mass function}
\acro{PN}{pseudo-noise}
\acro{PPM}{pulse position modulation}
\acro{PRake}{Partial Rake}
\acro{PSD}{power spectral density}
\acro{PSEP}{pairwise synchronization error probability}
\acro{PSK}{phase shift keying}
\acro{8-PSK}[$8$-PSK]{$8$-phase shift keying}
\acro{GMSK}{Gaussian minimum shift keying}
 
\acro{FSK}{frequency shift keying}

\acro{QAM}{quadrature amplitude modulation}
\acro{QPSK}{quadrature phase shift keying}
\acro{OQPSK/PM}{OQPSK with phase modulator }

\acro{RD}[RD]{raw data}
\acro{RDL}{"random data limit"}
\acro{ric}[RIC]{restricted isometry constant}
\acro{rict}[RICt]{restricted isometry constant threshold}
\acro{rip}[RIP]{restricted isometry property}
\acro{ROC}{receiver operating characteristic}
\acro{rq}[RQ]{Raleigh quotient}
\acro{RS}[RS]{Reed-Solomon}
\acro{RSC}[RSSC]{RS based source coding}
\acro{r.v.}{random variable}                               
\acro{R.V.}{random vector}

\acro{SA}[SA-Music]{subspace-augmented MUSIC with OSMP}
\acro{SCBSES}[SCBSES]{Source Compression Based Syndrome Encoding Scheme}
\acro{SCM}{sample covariance matrix}
\acro{SEP}{symbol error probability}
\acro{SG}[SG]{sparse-land Gaussian model}
\acro{SIMO}{single-input multiple-output}
\acro{SINR}{signal-to-interference plus noise ratio}
\acro{SIR}{signal-to-interference ratio}
\acro{SISO}{single-input single-output}
\acro{SMV}{single measurement vector}
\acro{SNR}[\textrm{SNR}]{signal-to-noise ratio} 
\acro{sp}[SP]{subspace pursuit}
\acro{SS}{spread spectrum}
\acro{SW}{sync word}

\acro{TH}{time-hopping}
\acro{ToA}{time-of-arrival}
\acro{TR}{transmitted-reference}
\acro{TW}{Tracy-Widom}
\acro{TWDT}{TW Distribution Tail}
\acro{TCM}{trellis coded modulation}

\acro{UAV}{unmanned aerial vehicle}
\acro{uric}[URIC]{upper restricted isometry constant}
\acro{urict}[URICt]{upper restricted isometry constant threshold}
\acro{UWB}{ultrawide band}
\acro{UWBcap}[UWB]{Ultrawide band}            
\acro{wuric}[UWRIC]{upper weak restricted isometry constant}
\acro{wurict}[UWRICt]{UWRIC threshold}

\acro{WiM}[WiM]{weigh-in-motion}
\acro{WLAN}{wireless local area network}

\acro{wm}[WM]{Wishart matrix}                               
\acroplural{wm}[WM]{Wishart matrices}
\acro{WMAN}{wireless metropolitan area network}
\acro{WPAN}{wireless personal area network}
\acro{wric}[WRIC]{weak restricted isometry constant}
\acro{wrict}[WRICt]{weak restricted isometry constant thresholds}
\acro{wrip}[WRIP]{weak restricted isometry property}
\acro{WSN}{wireless sensor network}                        
\acro{WSS}{wide-sense stationary}

\acro{sss}[SpaSoSEnc]{sparse source syndrome encoding}

\acro{VLC}{visible light communication}
\acro{RF}{radio frequency}
\acro{FSO}{free space optics}
\acro{IoST}{Internet of space things}
\end{acronym}

\title{CubeSat Communications: Recent Advances and Future Challenges}
\author{Nasir Saeed,~\IEEEmembership{Senior Member,~IEEE},  Ahmed Elzanaty, ~\IEEEmembership{Member,~IEEE}, Heba Almorad,~\IEEEmembership{Student Member,~IEEE},  Hayssam Dahrouj,~\IEEEmembership{Senior Member,~IEEE}, Tareq Y. Al-Naffouri,~\IEEEmembership{Senior Member,~IEEE}, Mohamed-Slim Alouini,~\IEEEmembership{Fellow,~IEEE}
\thanks{This work is supported by Office of Sponsored Research (OSR) at King Abdullah University of Science and Technology (KAUST). Hayssam Dahrouj would like to thank Effat University in Jeddah, Saudi Arabia, for funding the research reported in this paper through the Research and Consultancy Institute.

N. Saeed, A. Elzanaty, T. Y. Al-Naffouri and M.-S. Alouini are with the Computer Electrical and Mathematical Sciences \& Engineering (CEMSE) Division, KAUST, Thuwal, Makkah Province, Kingdom of Saudi Arabia, 23955-6900. H. Almorad and H. Dahrouj are with the department of Electrical and Computer Engineering at Effat University, Jeddah, Makkah Province, Kingdom of Saudi Arabia, 22332.}
}
\maketitle
\begin{abstract}
Given the increasing number of space-related applications, research in the emerging space industry is becoming more and more attractive. One compelling area of current space research is the design of miniaturized satellites, known as CubeSats, which are enticing because of their numerous applications and low design-and-deployment cost. The new paradigm of connected space through CubeSats makes possible a wide range of applications, such as Earth remote sensing, space exploration, and rural connectivity. CubeSats further provide a complementary connectivity solution to the pervasive Internet of Things (IoT) networks, leading to a globally connected cyber-physical system. This paper presents a holistic overview of various aspects of CubeSat missions and provides a thorough review of the topic from both academic and industrial perspectives. We further present recent advances in the area of CubeSat communications, with an emphasis on constellation-and-coverage issues, channel modeling, modulation and coding, and networking. Finally, we identify several future research directions for CubeSat communications, including Internet of space things, low-power long-range networks, and machine learning for CubeSat resource allocation.
\end{abstract}
\IEEEpeerreviewmaketitle

\section{Introduction}
The race for commercial dominance of the space industry is igniting, leading to an active new space economy. According to a report by Morgan Stanley, the expected revenue from the space industry will reach \$22 billion by 2024 and \$41 billion by 2029 \cite{s1}. Space businesses are growing especially fast with the development of small satellites because of the latter's relatively low deployment cost. Additionally, these small satellites are deployed in \ac{LEO}, thus providing low-latency communications \cite{Woellert2011}.
With the development of these small satellites, space technology is becoming cheaper, closer, and smaller, reviving the space industry by offering various new applications such as space observation, earth observation, and telecommunications. These provocative applications for small satellites are impelling top-tier companies like Google, SpaceX, OneWeb, and Facebook to investigate the use of these low-cost satellites, to provide earth monitoring, disaster prevention, and connectivity to Internet  of things (IoT) devices in remote areas instead of using the conventional LEO satellites.

In addition to industry growing interest, academic researchers have eagerly jumped into the development of small satellites. These satellites are often classified according to their weights, i.e., femto (less than $0.1$ kg), pico ($0.1$-$1$~kg), nano ($1$-$10$ kg), micro ($10$-$100$ kg), and mini ($100$-$1000$ kg) \cite{Woellert2011}. Among these, pico-satellites, also known as CubeSats, have emerged lately as the most popular ones. The CubeSat program was initiated at Stanford University in 1999 for the purpose of building a low-cost and low-weight satellite. Thereafter, the project's standards were defined {to build} these satellites in a cubic structure (hence the CubeSat terminology), with a mass of $1.33$ kg per unit, a cost of less than \$$1000$, low power consumption, and  off-the-shelf commercial components. { The cubic shape for these satellites was adopted because it provides sufficient surface area for solar power generation while providing better space-thermal stability. The basic unit for the CubeSat satellite was defined as a $1$U cube with dimensions of $10$ cm $\times$  $10$ cm $\times$ $10$ cm. Based on this unit, CubeSats vary in size from $1$U to $16$U \cite{isis}; so, for instance, $2$U CubeSats would have a mass of $2.66$ kg and dimensions of $10$ cm $\times$  $10$ cm $\times$ $20$ cm.
Taking advantage of the cubic structure, each face of the CubeSat consists of eight body-mounted solar cells or efficient solar panel wings. The solar panel wings considerably generate more power, i.e.,  $20$-$60$ W in full sunlight compared to eight mounted solar cells, which produce only $1$-$7$ W  \cite{Franco2018}.  However, the CubeSats are restricted to use a maximum transmission power of $1$ W ($30$ dBm) for establishing the communication link from CubeSat to the ground station, i.e., in the downlink direction. Nevertheless, the ground stations can use much higher transmission power, i.e., $100$ W ($50$ dBm), to send the data from the ground to the CubeSat, i.e., in the uplink direction \cite{Barbaric2018}.

\begin{table*}
\caption{List of a few well-known CubeSat missions and small satellite projects.}\label{table:missions}
 \centering
\begin{tabular}{|p{0.8cm}|p{0.75cm}|p{1.2cm}|p{3.8cm}|p{1.5cm}|p{2.2cm}|p{1.0cm}|p{2.8cm}| }
\hline
 \hline
    \textbf{Ref.}                          &\textbf{Launch Year}                        & \textbf{Mission Name}     &\textbf{Mission Type}    & \textbf{Size}   &\textbf{Frequency Band    } &\textbf{No of CubeSats}          &\textbf{Status}\\
       \hline\hline
       \cite{kipp}                    &2018                              &KIPP        &Providing global connectivity                    &3U       &Ku-Band         &2               &Operational\\
    \cite{radix}                        &2018                            &Radix        &Optical communication test     &6U        &Optical          &1            &Successfully completed\\
     \cite{gomix}        &2015                          &GOMX-3        &Aircraft signal acquisition     &3U        &X-Band           &1             &Successfully completed\\
     \cite{spire}                       &2018                      &Lemur-2          &Weather forecasting            &3U         & -           &100
     &Operational\\
     \cite{DICE}                    &2011                           &DICE            &Ionosphere Monitoring        &1.5U      &UHF-Band          &2             &Successfully completed\\
     \cite{Quaksat}                         &2003                        &QuakeSat          &Earthquakes forecasting     &3U       &UHF-Band            &1              &Successfully completed\\
     \cite{BUDIANU201514}                    &  -                          &OLFAR             &Low radiations analysis     & -          & VHF            &50-1000         &Under review\\
    \cite{POGHOSYAN201759}                   &2010                    &RAX               &Space weather forecasting    &3U       &S-Band             &2          &Successfully completed\\
    \cite{marco_2019}                               &2018                         &MarCO          &Relaying for deep space       &6U     &UHF and X-Band        &2           &Not Operational\\
    \cite{isara_2019}                   &2017                        &ISARA           &Bandwidth communication test      &3U      &Ka-Band                &1      &Operational\\
    \cite{aerocube_2019}  &2015   &AeroCube OCSD  & Optical communication speed test  &1.5U   &Optical              &2   &Successfully completed\\
     \cite{asteria_2019}                     &2017                        &ASTERIA               &Attitude control test   &6U                 &S-Band               &1   &Operational\\
     {  \cite{starlink}}                     &{  2019                       } &{  Starlink}               &{  Ubiquitous Internet  connectivity  } &{  Not a CubeSat}                 &{  X-band and Ku-band}               &{  42000} &{  Partially operational}\\
     {  \cite{oneweb} }                    &{  2019}                        &{  OneWeb} &{  Ubiquitous Internet  connectivity}   &{  Not a CubeSat}                 &{  Ku-band }              &{  650} & {  Partially operational}\\
     {  \cite{telesatleo} }                    &{  2019}                        &{  Telesat LEO} &{  Ubiquitous Internet  connectivity}   &{  Not a CubeSat}                 &{  Ka-band }              &{  200} & {  Partially operational}\\
     {  \cite{kuiper} }                    &{  2019}                        &{  Kuiper} &{  High-speed broadband services}   &{  Not a CubeSat}                 &{  Ka-band }              &{  3236} & {  Partially operational}\\
 \hline
 \hline
\end{tabular}
\end{table*}

Since their inception, CubeSats have risen to prominence {  as} they can perform many scientific experiments for educational and institutional purposes due to their tiny size \cite{Simons2015}. In fact, over a thousand different CubeSat missions  have been launched over the past 20 years \cite{kulu_2019}. These missions have fallen into four broad fields: communications, earth remote sensing, space tethering, and biology  \cite{inproceedings} \cite{article}. A few examples of these missions are included later in this survey to illustrate their capabilities.

Recently, both academia and the space industry have investigated the application of CubeSats to provide global connectivity to users around the world. This has led to the develoment of diverse projects and service offering in this field. For instance, KEPLER Communication launched their KIPP CubeSats in 2018 to provide connectivity to users at the North and South Poles. KIPP was the first Ku-band 3U CubeSat mission  to offer a 40 Mbps data rate with a $60$-cm diameter very-small-aperture terminal (VSAT) \cite{kipp}.  Analytical Space launched its Radix mission to enable high-speed downlink communication using optical links.  Radix  consisted of 6U CubeSats for the primary purpose of testing laser capabilities in downlink communications \cite{radix}. It began transmitting data soon after it was deployed and continued for six months. 
Another well-known CubeSat mission was GOMX-3, supported by the European space agency (ESA). GOMX-3 consisted of 3U CubeSats, and was successful in acquiring signals from the aircrafts worldwide \cite{gomix}.

Besides their use in communications, CubeSats were employed for a number of missions dedicated to the scientific understanding and prediction of the earth's environment. Weather prediction, climate change, and disaster monitoring are the most common applications for earth science missions. For example, Lemur-2 was a LEO-constellation  CubeSat mission from Spire \cite{spire} in which the satellites consist of two payloads-one for weather prediction and one for ship tracking.

 Dynamic Ionosphere CubeSat Experiment (DICE) was another well-known mission led by Utah State University to monitor the Earth's ionosphere \cite{DICE}. DICE CubeSats measured in-situ plasma densities using two Langmuir probes for geospace storm-time features identification. Launched in 2003, the QuakeSat (3U CubeSat) mission forecasted earthquakes. It carried a magnetometer housed in a $60$-cm telescope for the purpose of scanning and collecting global changes and fluctuations in extremely-low-frequency-electromagnetic (ELF) waves, which are believed to precede seismic activity \cite{Quaksat}.

CubeSats are also popular in universe-exploration and space-science missions, which aim to expand the scientific knowledge in astronomy, heliophysics (space weather), and planetary science. Analyzing cosmic background radiation below $30$ MHz is exceptionally challenging, since it requires space-based large-aperture radio telescopes, which are quite sensitive to ultra-long waves in space. Therefore, a  distributed system consisting of a swarm of $50$ or more CubeSats in lunar orbit called orbiting low-frequency antennas for radio astronomy (OLFAR) has been used to analyze cosmic radiation \cite{BUDIANU201514}. Another vital space-exploration mission consisting of $3$U CubeSats was the Radio Aurora Explorer (RAX), launched to investigate the formation and distribution of natural ionospheric plasma turbulence,  for the purpose of enhancing the space-weather forecasting and minimizing  damage to satellites and spacecraft technologies \cite{POGHOSYAN201759}. Another remarkable space-science CubeSat mission was Mars Cube One (MarCO), launced in 2018 by NASA consisting of $6$U CubeSats. MarCO was the first flying CubeSat mission to deep space that supported  relaying telecommunications to Mars \cite{marco_2019}.

{   Another popular application for small satellites (not CubeSats) is providing ubiquitous Internet connectivity for consumers and IoT devices. For instance, the Starlink project by SpaceX will deploy thousands of satellites to satisfy consumer demands for high-speed and reliable Internet around the globe, especially in the regions where there is no connectivity, or when other solutions are too expensive \cite{starlink}. A direct competitor with SpaceX, OneWeb will provide global connectivity solutions with a blanket of small LEO satellites \cite{oneweb}. In its first phase,  OneWeb will launch $650$ satellites; the next step will launch $400$ more satellites to enhance the global coverage. Both of these projects are still in their early stages; however, they are expected to become the mainstream Internet providers from space. Other well-known ambitious projects include Telesat LEO \cite{telesatleo} and Kuiper \cite{kuiper}. All the aforementioned CubeSat missions and small satellite projects are summarized in Table~\ref{table:missions}; please also see references \cite{kulu_2019, BOUWMEESTER2010854, Swartwout2013, Mukherjee2013, Villela2019} for a complete list of academic and industrial CubeSat missions.
}

\subsection{Related Review Articles}\label{sec: relatedsurveys}
The CubeSat literature includes several review articles, e.g., \cite{BOUWMEESTER2010854, Swartwout2013, Mukherjee2013, Villela2019, Radhakrishnan2016, Schaire2016, Rahmat2017, Sweeting2018, Franco2018,  Gregorio2018}, the summary of which can be found in Table~\ref{Tablerelatedsurveys}. The first comprehensive history of CubeSat missions through 2009 is presented in \cite{BOUWMEESTER2010854}. The literature on these missions is further extended  to the year 2013 in \cite{Swartwout2013}, in which the statistics of CubeSats missions are briefly enumerated. These statistics  include the number of missions launched, launch failures, operational missions, non-operational missions, and failed missions from the year 2000 to 2012.  Reference  \cite{Swartwout2013} further details the ratios of these missions in different parts of the world until 2012. Mukherjee \textit{et al.}  offers a comprehensive survey of space networks and interplanetary Internet, as well as presents the concept of delay-tolerant networking for deep-space networks \cite{Mukherjee2013}.  However, the focus of their paper is not on CubeSats, but rather on the networking component of deep space networks.
In \cite{Villela2019}, Villela \textit{et al.} extend the data on CubeSat missions through 2018, detailing the number of countries involved in CubeSat research, the success rate of CubeSat missions, and predicting that a thousand CubeSats will be launched in 2021. A brief survey of  CubeSats inter-satellite communications is presented in \cite{Radhakrishnan2016}, which  focuses on enabling inter-satellite communications by examining the physical, network, and medium-access control layers of the open-system interconnection (OSI) model for small satellites.
\begin{table}[t]
\footnotesize
\centering
\caption{Comparison of this paper with the existing surveys.}
\label{Tablerelatedsurveys}
\begin{tabular}{|p{2.2cm}|p{0.5cm}|p{4.9cm}|}
\hline
\hline
 \textbf{Ref.}              & \textbf{Year}         & \textbf{Area of Focus}  \\ \hline\hline
Bouwmeester \emph{et al.} \cite{BOUWMEESTER2010854}                     & 2010  & History of CubeSat missions from 2000 to 2009 \\ \hline
Michael  \textit{et al.} \cite{Swartwout2013}                      & 2013          & History and statistics on CubeSats missions from 2000 to 2012 \\ \hline
Joyeeta \textit{et al.} \cite{Mukherjee2013}                      & 2013           & Deep space networks and interplanetary Internet       \\ \hline
Thyrso \textit{et al.} \cite{Villela2019}                      & 2016           & History and statistics on CubeSats missions from 2000 to 2016          \\ \hline
Radhika \textit{et al.} \cite{Radhakrishnan2016}                      & 2016   & Inter-satellite communications for CubeSats \\ \hline
Scott \textit{et al.} \cite{Schaire2016}                      & 2016       & NASA's near earth network and space network for CubeSats      \\ \hline
Yahya \textit{et al.} \cite{Rahmat2017}                      & 2017       & Antenna designing for CubeSats        \\ \hline
Martin \cite{Sweeting2018}                      & 2018       & History, statistics, and applications of CubeSats missions        \\ \hline

Franco \textit{et al.} \cite{Franco2018}                      & 2018       &  Small satellite missions,  antennas design, and networking        \\ \hline

Anna \textit{et al.} \cite{Gregorio2018}                      & 2018       &  Hardware challenges of CubeSat missions        \\ \hline

This paper                      & 2019       &  Coverage and constellation issues, channel modeling, modulation and coding, networking and upper layer issues, and future research challenges for CubeSats       \\
\hline
\hline
\end{tabular}
\end{table}

In \cite{Schaire2016}, Schaire \textit{et al.} summarize the support, services, and future plans offered to the emerging CubeSat market by NASA's Near Earth Network (NEN), Space Network (SN), and Space Communication and Navigation Network (SCaN). The authors also discuss the capabilities of the NEN and SN, illustrating the maximum achievable data rates and data volumes for different orbit altitudes and slant ranges.
The literature on the development of CubeSat antennas is summarized in \cite{Rahmat2017}, which discusses the types of antennas used for CubeSats, including horn, patch, dipole, reflector, and membrane antennas. Reference \cite{Sweeting2018} surveys  the literature on evolution, constraints, policies,  and applications of small satellites. Davoli \textit{et al.} present an overview of the different physical aspects of small satellites, which include hardware components, antennas design, and networking \cite{Franco2018}.   Gregorio \textit{et al.} describe the hardware-based challenges facing CubeSat missions, including miniaturization, power control, and configuration \cite{Gregorio2018}.

In summary, most of the above surveys focus on the quantitative details of CubeSat missions, e.g., the number of missions, the number of CubeSats, launching dates, participating countries, and mission targets \cite{BOUWMEESTER2010854, Swartwout2013, Sweeting2018, Villela2019}. However, only a few surveys discuss the communications features of CubeSats, e.g., inter-satellite networking \cite{Mukherjee2013}, antenna design \cite{Franco2018}, and delay-tolerant networking \cite{Schaire2016}.

\subsection{Contributions of this Paper}
Despite the plethora of works on CubeSats, as highlighted in Section \ref{sec: relatedsurveys}, to the best of ours knowledge, there is no consolidated article that provides a comprehensive survey of CubeSat communication system.  More importantly, the existing surveys do not connect the effects of important technical considerations, such as modulation, coding, networking, constellation design, and cost constraints of CubeSat communication system.

In this paper, we offer an overview of various features of CubeSats that significantly influence the performance of their communication system. Also, we answer some research questions that can provide insights about CubeSat communications, for example, how the constellation type, the CubeSat altitude, and the mission targets affect link performance. Moreover, we address the question of designing efficient communication systems that take into consideration these features.

Initially, we provide an overview of CubeSat constellation design. The choice of a particular constellation type depends on the goal of the mission. For instance, if the aim is to provide global communication coverage, a larger number of satellites is required. In this case, we thereby discuss the number of orbital planes and satellites required to achieve ubiquitous coverage. For remote sensing applications, a smaller number of CubeSats is typically needed. We also discuss the question of how to extend  data coverage using CubeSats for communications in rural and remote areas of the world, an {  emerging topic} in CubeSat design.

Next, we examine the impact of satellite geometry on the communication channel model. We describe the evolution of the statistical channel models adopted for satellite communications from land mobile satellites (LMS) to CubeSats. Various channel models from the literature are compared according to their relevance to CubeSats.

Then, we examine the CubeSat link budget with respect to satellite geometry, operating frequency, and channel modeling. The parameters of the link budget depend heavily on the characteristics of CubeSats, for instance, limited power and antenna gain. Subsequently, we introduce the modulation and coding techniques usually adopted in CubeSat research. Moreover, we show that how these schemes incorporate the link budget and elevation angle (angle between the satellite and earth) to provide reliable communications. Also, we compare these techniques and present the recommendations suggested by the \ac{CCSDS}.

Later, we briefly discuss the physical networking of CubeSats based on different communication technologies such as, \ac{RF} and \ac{FSO}, including laser and \ac{VLC}. We also discuss the routing protocols used for satellite-to-ground and inter-satellite links. Lastly, we anticipate future research directions and major open issues in CubeSat research, for instance, heterogeneous CubeSats-6G networks, software-defined networking, \ac{IoST}, hybrid architectures, ubiquitous coverage, and machine learning.

The contributions of this paper are summarized as follows:
\begin{enumerate}
\item We provide a comprehensive survey of CubeSat communications, which we envision to enable \ac{IoST}.

\item We discuss the most hotly-debated topic in the domain, the extended coverage using CubeSats for communications in rural and remote areas of the world.

\item We survey the technical dimensions of CubeSats communications, including channel modeling, modulation and coding schemes, networking, constellation design, and  coverage issues.

\item We present several future research directions for CubeSats and their application to earth remote sensing, space sensing, and global communications.
\end{enumerate}

\subsection{Organization of the Paper}
This paper is organized as follows: Section II presents an overview of CubeSat constellation designs and coverage concepts. Section III covers various communication links and channel modeling techniques for CubeSats. Section IV discusses the link budget calculation for CubeSat communications.
Sections V and VI explain modulation and coding methods and medium access control (MAC) protocols, respectively. Section VII and VIII discuss the networking and application layer protocols for CubeSats, respectively. In Section IX, we focus on future research challenges for CubeSat communications. Finally, Section X summarizes and concludes the survey.

\section{CubeSat Constellations and Coverage}
The coverage of any CubeSat mission depends on different parameters such as the number of satellites, the number of orbital planes, the elevation angle, the inclination,  the altitude, the orbital plane spacing, and the eccentricity of the orbit. To date, CubeSats have been deployed in LEO orbits because of their lower cost and lower implementation complexity. However, the footprint of LEO satellites is much smaller than those of medium earth-orbit (MEO) and geostationary earth-orbit (GEO) satellites. To put this in perspective, more than $100$ LEO satellites are required for global coverage, compared to fewer than ten MEO satellites \cite{Lazreg2016}. {  To elaborate on this, first we will discuss the beam coverage of CubeSats before moving into the different types of satellite constellations.}
\begin{figure}
\begin{center}
\includegraphics[width=0.80\columnwidth,clip]{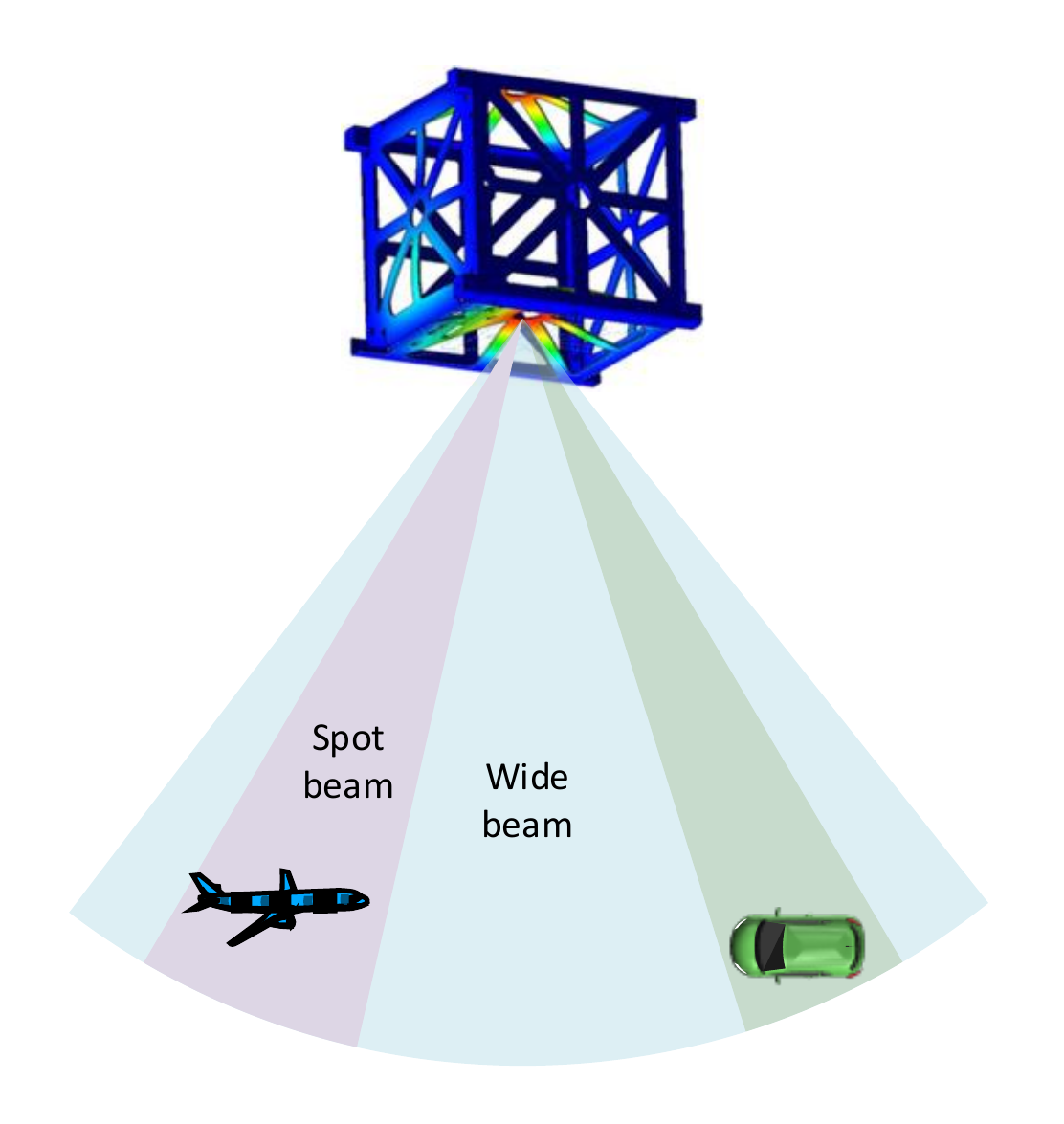}
 \caption{Hybrid beam scheme for CubeSats in LEO.}
 \label{fig: hybrid}
\end{center}
\end{figure}
{  
\subsection{Beam Coverage}
The beam coverage of an individual CubeSat depends mainly on the satellite's orbital altitude and type of antenna used. CubeSat antennas must have low loss, spherical coverage, high reliability, and a compact size. Full spherical coverage is typically achieved with multiple antennas. Typical CubeSat antennas for telemetry and telecommand include microstrip patches, monopoles, turnstiles, and helical antennas. Once a CubeSat is stable in orbit, it requires high-speed communications with the ground station, which in turn requires high-gain and compact-size antennas. These high-gain antennas must be able to point their beams accurately. The ideal coverage pattern for CubeSats operating in the S- and X-bands is an isoflux pattern which takes into account free-space propagation losses. At higher frequency bands such as the X-band, high-gain horn antennas with beam steering capabilities can be used to enable wider bandwidth and accurate pointing towards the ground station. The small size of CubeSats limits the use of large-sized antennas, and therefore, various efforts have been made to design small-sized high-gain antennas. A parabolic reflector of size $0.5$ m, designed in \cite{sstl}  for $1.5$U CubeSat, is compatible with NASA's deep-space networks. Consequently, a folded panel reflectarray was proposed in \cite{Qin2015} with better stowage efficiency, low cost, and beam pointing capability when compared with the conventional reflectors in \cite{sstl}. Interested readers are referred to \cite{Gao2018} and references therein for various CubeSat antennas with different beam patterns.

Seamless global coverage can be attained by adjusting the satellite antennas' beam patterns. For instance, in \cite{Su2019}, Su \textit{et al.} present a hybrid wide- and spot-beam schemes for LEO satellites. In the hybrid approach, a wide beam operating at a low-frequency band is used for a large coverage area, while spot beams at high-frequency bands facilitates high-speed data access (see Fig. \ref{fig: hybrid}).
}

\subsection{Constellation Design}
Besides orbital altitude, satellite constellation design is a major factor affecting the coverage of CubeSat missions \cite{Marinan2013}. {  A constellation is a group of satellites that coordinates its operations so as to provide global or near-global coverage. Satellite constellations typically consist of complementary orbital planes.} There are mainly three constellation designs for providing global coverage:
\begin{itemize}
 \item Walker constellations: Walker constellation design is symmetric, in other words, all the satellites have the same inclination and latitude. The parameters of Walker constellations are defined as inclination $i$, number of satellites $N_s$, number of equally spaced orbital planes $N_p$, and relative phase difference between the planes $\Delta\phi$. Based on these parameters, each orbital plane has $n =\frac{N_s}{N_p}$ number of satellites, where the inclination of all the planes is same (see Fig.~\ref{fig:walker}).
Latitudinal zones which are beyond the inclination angle of the orbital planes may not have any coverage in Walker constellations.
 \begin{figure}[h]
\begin{center}
\includegraphics[width=0.70\columnwidth]{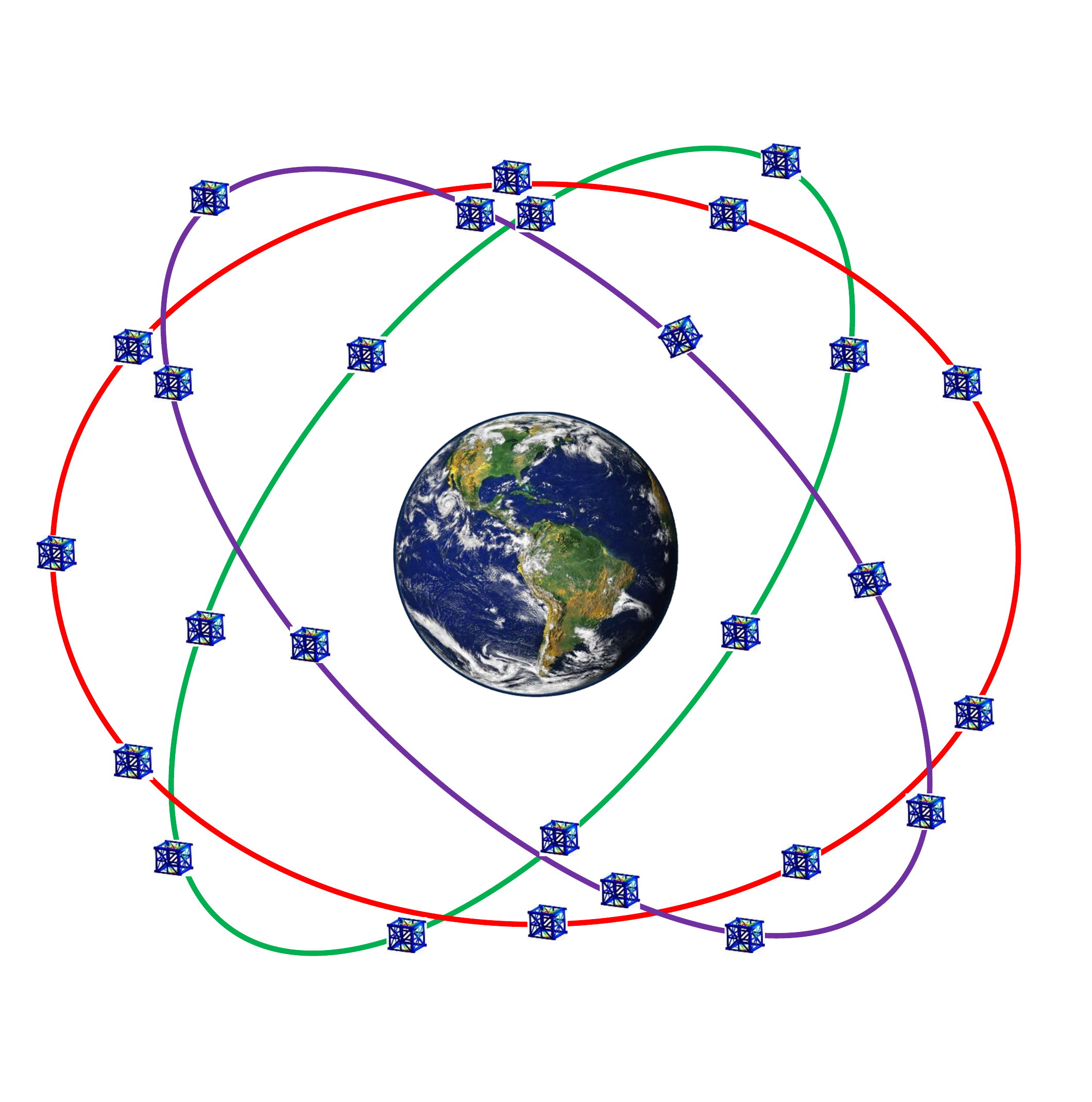}
 \caption{Illustration of Walker constellation for GALILEO.}
 \label{fig:walker}
\end{center}
\end{figure}

To design and analyze a CubeSat constellation for a longitudinal global coverage, the minimum number of CubeSats $n$ per orbital plane, and the minimum number of planes $N_p$ required for a circular orbit can be determined as,
\begin{equation}\label{eq: ns}
n = \left\lceil{\frac {360}{2 \theta}}\right\rceil,
\end{equation}
and
\begin{equation}\label{eq: np}
N_p =\left \lceil{\frac {360}{4 \theta}}\right\rceil,
\end{equation}
respectively, where $\lceil{.}\rceil$ is the {  ceiling function} and $\theta$ is the Earth central angle of coverage. By using the law of sine, the Earth central angle is obtained as follows \cite{AKYILDIZ2019166},
\begin{equation}
\theta = \arcsin \left( \frac{\rho\,   \sin (90+ \phi)}{h+R_E}\right),
\end{equation}\label{eq: th}
\noindent where $R_E$ is the Earth's radius, $h$ is the orbital altitude of the CubeSat, $\phi$ is the elevation angle, and $\rho$ is the slant range (see Fig. \ref{fig:coverage}). The slant range $\rho$ can be determined by the law of cosines as,
\begin{equation}
\rho^2 - 2 R_E\, \rho\,\cos{(90+ \phi)}  = (R_{E}+h)^2 - R_{E}^{2}\,.
\end{equation}\label{eq: th}
To illustrate the effect of the orbital altitude and elevation angle on $n$ and $N_p$, we plot (\ref{eq: ns}) and (\ref{eq: np}) in Figs. \ref{fig: planes} and \ref{fig: noofCubeSats}, respectively. As shown in Figs. \ref{fig: planes} and \ref{fig: noofCubeSats}, the CubeSat requirement per orbital plane and the number of orbital planes increase with increasing the elevation angle and reducing the altitude. This is due to the direct relation between the Earth central angle $\rho$, the elevation angle $\phi$, and the orbital altitude $h$. Furthermore, for a fixed altitude, increasing the elevation angle from $5^\circ$ to $25^\circ$ leads to increasing the number of planes and the number of  CubeSats per plane. However, reversing this scenario, for instance, keeping the elevation angle fixed and increasing the altitude from $500$ km to $900$ km, reduces the number of planes and the number of CubeSats per plane due to better coverage at higher altitudes.
 \begin{figure}
\begin{center}
\includegraphics[width=1\columnwidth]{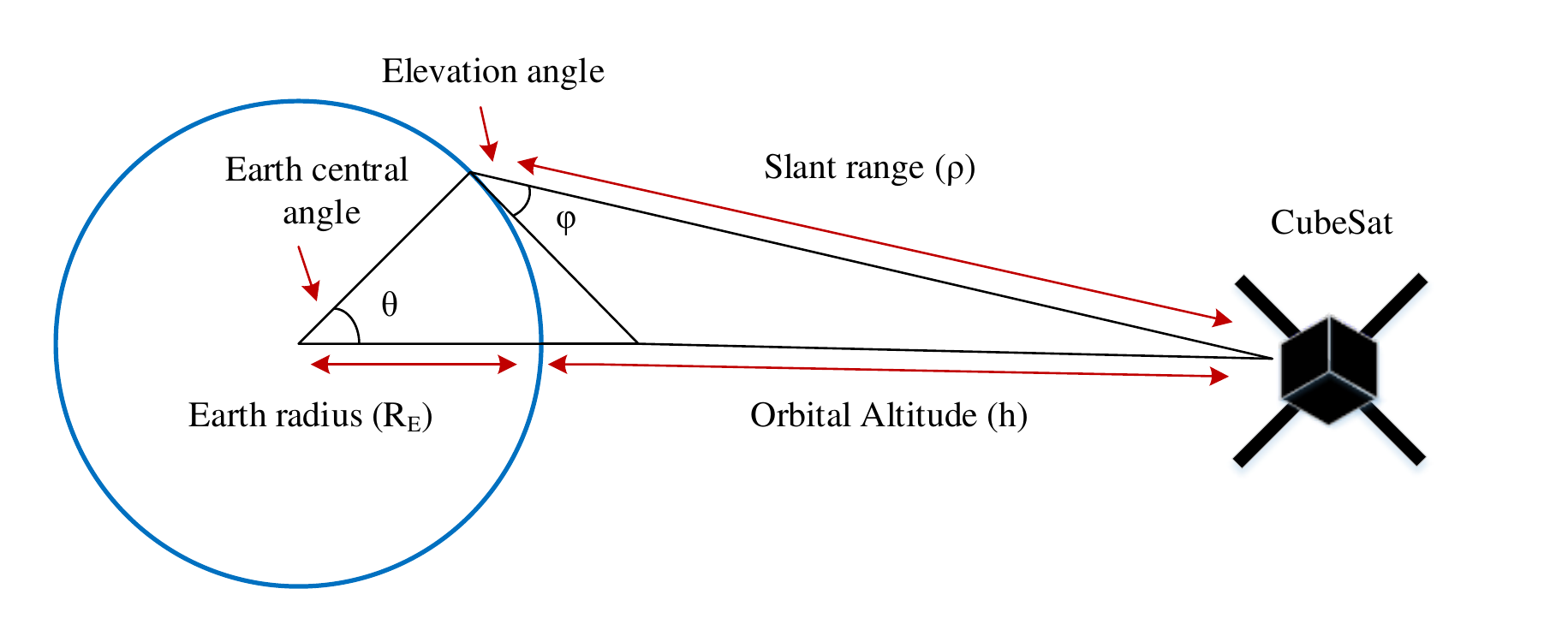}
 \caption{Coverage geometry for CubeSats.}
 \label{fig:coverage}
\end{center}
\end{figure}
\begin{figure}[h]
\centering
\includegraphics[width=0.8\columnwidth]{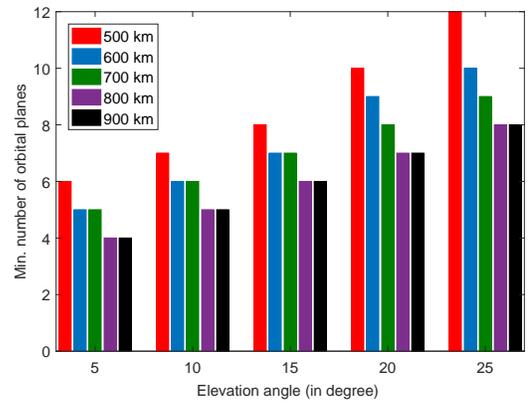}
\caption{Elevation angle versus number of orbital planes required.}\label{fig: planes}
\end{figure}
\begin{figure}[h]
\begin{center}
\includegraphics[width=0.8\columnwidth]{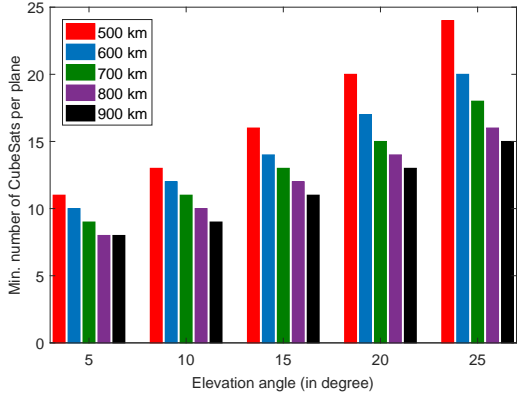}
 \caption{Elevation angle versus number of CubeSats per plane.}
 \label{fig: noofCubeSats}
\end{center}
\end{figure}
\begin{figure}
\begin{center}
\includegraphics[width=0.38\columnwidth]{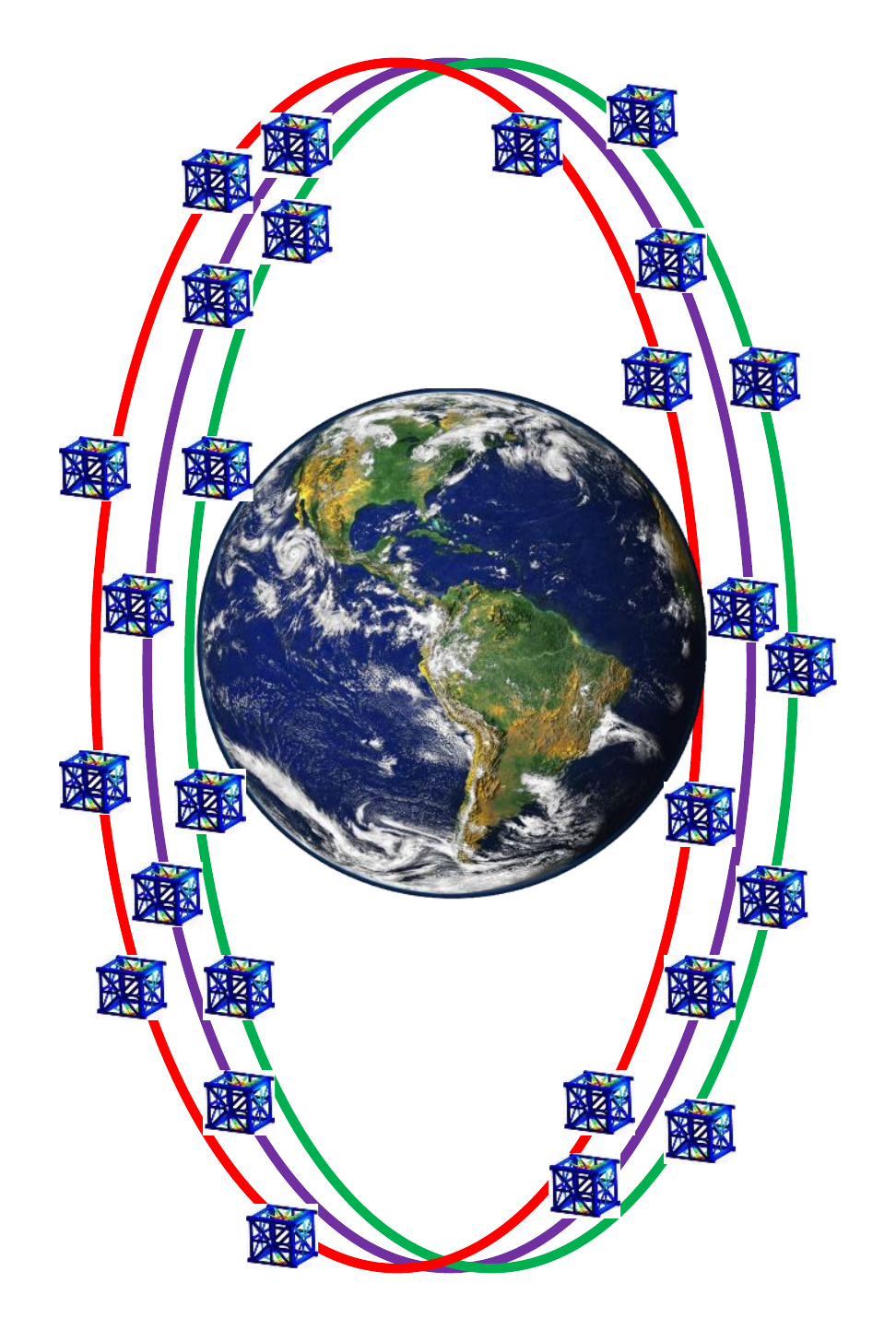}
 \caption{Illustration of a polar-inclined, street-of-coverage constellation.}
 \label{fig:soc}
\end{center}
\end{figure}
\begin{figure*}[htb!]
\begin{center}
\includegraphics[width=2\columnwidth]{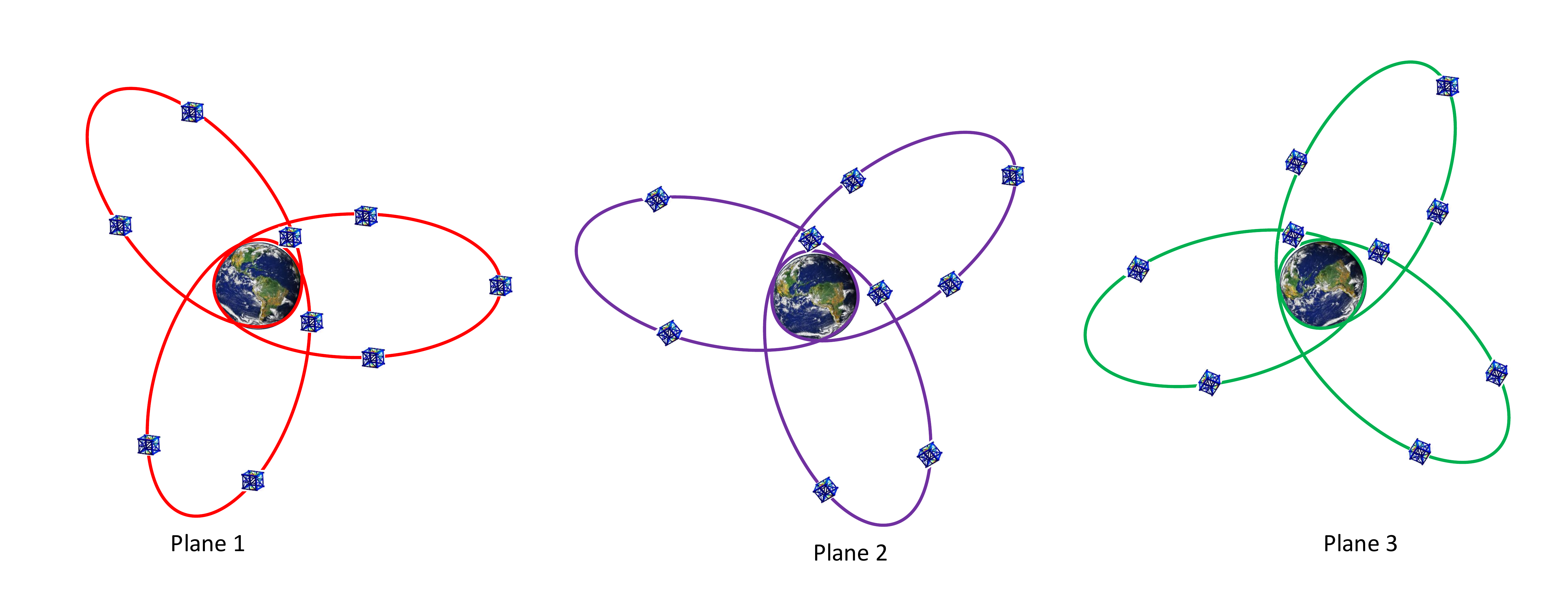}
 \caption{Illustration of a Flower constellation in three different orbital planes.}
 \label{fig: flower}
\end{center}
\end{figure*}
 \item Street-of-coverage constellations: These constellations consist of non-uniformly distributed, polar-inclined orbital planes. The separation between the orbital planes and their phase difference is designed in such a way that adjacent planes overlap with the coverage region so as to provide global coverage. A major issue with these constellations is that Earth coverage is not uniform, with the highest coverage at polar regions and lowest coverage at equatorial regions, as shown in Fig. \ref{fig:soc}. Missions that emphasize coverage at the equatorial region require orbital planes to be equally spaced by  $180^{\circ}$; however, such a constellation design requires longer deployment time and multiple launch sites \cite{Su2017}.

\item Flower constellations: The idea of a Flower constellation was first introduced in 2003 \cite{McManus2016}. Flower constellations consist of satellites following the same closed-loop trajectory in a rotating frame of reference \cite{Avendano2012}. The Earth-centered-Earth-fixed reference frame is used where all the satellites are synchronized and coordinated with the rotation of the Earth. The orbital planes in Flower constellations satisfy the following condition \cite{Avendano2012}
\begin{equation}
P_p T_p = P_d T_d,
\end{equation}
where $P_p$ and $P_d$ are co-prime integers, $T_d$ is the time period of the rotating reference frame, and $T_p$ is the rational multiple of $T_d$. Also, the semi-major axis, orbit inclination, perigee argument, and eccentricity of the orbits are the same. Furthermore, the mean anomaly $A_i$ of the $i$-th satellite satisfies $P_p Q_i = -P_d A_i \mod(2 \pi)$,
where $Q_i$ is the right ascension of the ascending node.
Fig. \ref{fig: flower} shows an example of such constellation design, where a group of three orbits on the same orbital plane is used with the same inclination, eccentricity, and semi-major axis. Flower constellations provide some interesting orbital mechanics for flying formation and can support both regional and global area services.
\end{itemize}

Satellite constellation design can also be configured for a specific mission. For example, in \cite{nag2016}, Nag \textit{et al.} proposed that the Walker constellation can be used to provide the air-traffic surveillance in the Alaska region. Two perpendicular orbital planes with eight satellites in each plane were used to provide 99\% coverage in Alaska.

\subsection{Swarm of CubeSats}
Besides their use in constellation design, there is an on-going interest in the use of swarms or clusters of small satellites in missions. A satellite swarm can certainly improve mission' coverage, both in space and on Earth. The concept of the satellite swarm was introduced by the U.S. Defense Advanced Research Projects Agency (DARPA)  with the  F6  system (Future, Fast, Flexible, Fractionated, and Free-Flying) \cite{DARPA}. In the F6 system, a traditional satellite was inserted among the cluster of sub-satellites where the resources were shared among the sub-satellites using inter-satellite communication. Although the F6 system was the first step towards the satellite swarm, it was canceled after two attempts, since an integrator needed to pull the system together was missing \cite{DARPA2}. Inter-satellite communications and flight formation are the major concerns for the satellite swarms \cite{Alvarez2016}; hence, efforts have been made to optimize satellite-to-satellite coverage for  inter-satellite links. In \cite{Lluch2014}, Lluch \textit{et al.} optimized the satellite's  orbital parameters to provide maximum coverage with six LEO satellites; thus, an increase in the inter-satellite link improved the coverage of the mission.  Danil \textit{et al.} proposed a decentralized differential drag-based control approach for the cluster formation of $3$U CubeSats \cite{IVANOV2019646}. In the absence of a control strategy, the satellites in a cluster move apart from each other in the orbital plane. Therefore, it is important to model the aerodynamic drag force and reduce the relative drift between the satellites to zero. Recently, Cornell University launched $105$ tiny-sized satellite swarm, also called ChipSats in the KickSat-2 mission, which  successfully demonstrated that forming a swarm of small free-flying satellites is possible \cite{chipsat}. Not only do these tiny satellites further reduce the costs, they also improve coverage in both space and on Earth. In  \cite{Palma2018}, Palma \textit{et al.} proposed a free-flying satellite swarm to provide connectivity to the IoT networks in the Arctic region. Three different CubeSat orbital configurations  were considered, along with three CubeSats and four ground stations. They have shown that free-flying swarms of CubeSats achieve overheads below $27$\% and are therefore good candidates to support rural IoT networks.

{  
\subsection{Lessons Learned}
In this section, we discussed the coverage issues with CubeSat communications. First, we introduced beam-coverage, which depends mainly on the CubeSat's orbital altitude and type of antenna used. CubeSat antennas should be small in size and operate at low power due to the limited amount of on-board power. Different antennas are used depending on the size of the CubeSat and mission type. For instance, folded-panel reflect-array has better beam-pointing capabilities compared to conventional reflectors. Also, hybrid beams can be used to provide high-speed links and better coverage at the same time.

The second part of this section focused on the coverage of various constellation designs. Three different constellation designs, including Walker, street-of-coverage, and Flower, were briefly discussed. Each constellation has its pros and cons, for example, Walker constellations are easy to implement, but they have more coverage at polar regions than in equatorial areas. Similarly, Flower constellations can provide better coverage but are harder to implement. Constellations should be designed according to their applications, for instance, for covering a specific area on Earth's surface or in space.

Lastly, we surveyed the research on CubeSats swarms, which can improve the coverage both on the Earth's surface and in space. Significant challenges for CubeSats swarms include inter-satellite communication and flight formation. Inter-satellite communication can be accomplished either using RF or optical waves. RF waves provide better coverage at the cost of low data-rate links, while optical waves provide high-speed links but require accurate pointing and acquisition methods. Similarly, the flying formation for CubeSats can be designed by modeling the aerodynamic drag force and reducing the relative drift between the satellites.

}

\section{Communication Links and Channel Modeling}
{  
	One significant issue faced by CubeSat missions is the lack of a standardized communication channel model. Although the \ac{CCSDS} specifies some international standards, there are various obstacles for receiving telecommand signals in CubeSats using these standards. Mainly, these obstacles are due to the error correction and detection codes used in CCSDS standards. Here, we propose various channel models based on the CCSDS standards for CubeSat communications that depend mainly on the communication link. CubeSat communication links can be divided  into two types:  CubeSat-to-Ground (C2G) and CubeSat-to-CubeSat (C2C) links, (see Fig.~\ref{fig: iost}). 	In this section,  we discuss the various types of links in CubeSat communications and their corresponding  statistical channel models.
	\begin{figure}[h]
		\begin{center}
			\includegraphics[width=1\columnwidth]{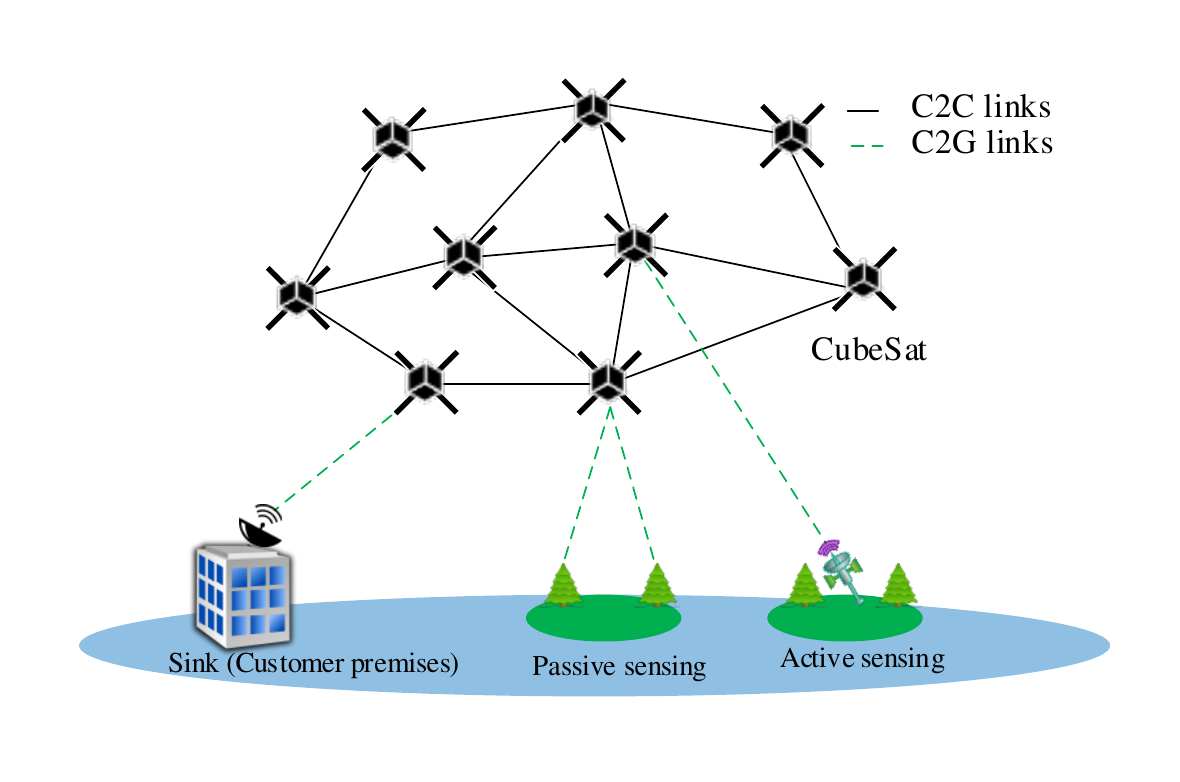}
			\caption{Networking architecture for CubeSats \cite{saeed2019}.}
			\label{fig: iost}
		\end{center}
	\end{figure}
	
	
	\subsection{CubeSat-to-Ground Communications}
	CubeSats employ various communication technologies to establish a C2G link, namely, VLC, laser, and RF. In \cite{Nakajima2012}, Nakajima \textit{et al.} used VLC-based micro-satellites for space-to-earth links, achieving a data rate of $9.6$ kbps under perfect alignment. The satellite altitude in \cite{Nakajima2012} is $400$ km with $40$ km footprint on Earth. On the other hand, laser communication have grown in importance, because they provide a large bandwidth, a license-free spectrum, a high data rate, less power, and low mass requirements. A laser communication system was used in Small Optical Transponder (SOTA) mission in 2014, which was able to achieve the data rate of $10$ Mbps for the downlink \cite{Alberto2017}. Lasers were also used in the Aerocube OCSD mission to demonstrate C2G links, providing high data rate and near-zero latency \cite{Richard2018}. Recently,  a laser based C2G link from a LEO $1.5$U CubeSat at a 450 km altitude to an optical ground station was established \cite{RoseRowen:19}. This communication link achieved a data rate of up to $100$ Mbps with bit error rates near $10^{-6}$. Since, pointing and acquisition are major problems for free-space optical communications,  a hybrid RF-and-optical approach is introduced in \cite{Welle2017}, where CubeSats are used as relay satellites between the GEO satellites and the ground station using both RF and optical links. Several research projects are initiated by the National Institute of Information and Communications Technology (NICT) Japan for using lasers in C2G links \cite{Kolev2018}.

Although the use of lasers can lead to high-speed C2G links, the atmosphere introduces following losses to the free-space optical communication:
\begin{itemize}
	\item Absorption and scattering losses resulting from the presence of various gas molecules and aerosols in the Earth's atmosphere.  The attenuation due to various weather conditions, e.g., aerosol particles, fog, and haze, depends on the frequency. Details regarding the prediction of these losses can be found in several databases \cite{RothGamaGold:87,KneShe:80,SmiAcc:93}.
	\item Atmospheric-turbulence, which is a random phenomenon, caused by variations in the temperature and pressure of the atmosphere along the propagation path.
	\item Beam divergence loss due to  beam diffraction near the receiver.
	\item Background noise from the Sun and other stars.
	\item Pointing loss due to satellite vibration or imperfect tracking-and-stabilization mechanisms.
\end{itemize}
These factors can cause a severe degradation in performance if they are not well compensated for. Some statistical models for optical channels along with several approaches to counterbalance  atmospheric effects have been reviewed in \cite{KauKad:17}; however, these models are still in the early research phase. Currently, RF is widely used to establish C2G links. For instance, the DICE mission achieved a data rate of $3$ Mbps operating in the UHF band \cite{DICE}. In \cite{Ennis2018}, the authors examined the  C2G link for the Tianwang-1 mission, which provided $125$ Kbps of maximum data rate. In the following, we focus more on the channel models for RF-based C2G communication.   Most of these models consider land-mobile-satellite (LMS) communication systems, in which mobile devices on Earth communicate with the CubeSats in LEO.
}

{   The major differences between the ground communication channel and the LMS channel are the limited power available on the CubeSats and the differing geometry of the CubeSat-to-ground link. Some of the LMS systems can only be accessed with  low-elevation angles. Also, the LMS channel is affected by additional impediments, such as the tropospheric and ionospheric losses, especially at low-elevation angles. Besides, the CubeSats' low power requires a strong line of sight, which makes establishing the link at low elevation angles particularly challenging.


The random variations that the signal envelope experiences are due to three main phenomena: multipath fading, line-of-sight (LOS) shadowing, and multiplicative shadow fading. Multipath fading arises from a combination of scattered non-LOS (NLOS) components along with a possible LOS ray, leading to rapid small-scale fluctuations.  LOS shadowing results from the partial blockage of the LOS by large-size objects, such as terrains and buildings, leading to large-scale, slow fluctuations.  Multiplicative shadow fading is responsible for the random variations in the power of both  LOS multipath components. Multipath fading is modeled using well-known statistical distributions such as Rayleigh and Rice distributions, while the shadowing is modeled using log-normal or Nakagami distributions.

The LMS channels can be broadly categorized into static and dynamic channel models.
 The following are the major statistical channel models that can be used for CubeSat communications (also summarized in Table \ref{table:lms}):

\subsubsection{Static Models}

The static LMS channel models mainly consider the LOS direct path, LOS diffused path, and the multipath \cite{Arapoglou2011}.  In static channel models, the distribution of the signal envelope can be modeled by a single distribution that does not change with time. Hence, it makes sense to describe the following static channels.

\paragraph*{Loo's Model \cite{Loo1985}}
It is one of the famous statistical channel models used for LMS systems. Loo's model assumes that the LOS component undergoes a log-normal shadowing, while the multi-path signals are  Rayleigh distributed \cite{Loo1985}.  First, the distribution  of  the signal envelope conditioned on  the LOS component $a$ is derived as
\begin{equation}
f(r|a) =  \frac{r}{\sigma^2_m  }
\exp\left({-\frac{r^2 + a^2}{2 \sigma_m^2}}\right)I_0\left(\frac{r a}{\sigma_m^2}\right)
\end{equation}
where $\sigma_m^2$ is the variance of the multipath, and $I_0(\cdot)$ is the zero-order Bessel function.
Then, the PDF of the envelope can be found by averaging the conditioned distribution over the random variable representing the LOS shadowing  as
\begin{equation}
f(r)  =  \int_{0}^\infty f(r|a) f_{\text{s}}(a)\,\textrm{d}a,
\end{equation}
where

\begin{equation}\label{eq: logpdf}
f_{\text{s}}(a) = \frac{1}{ a\sqrt{2 \pi \sigma_s^2 }}\exp\left(-\frac{(\log(a)-\mu_s)^2}{2\sigma_s^2}\right),  \quad a > 0,
\end{equation}
$\mu_s$ and $\sigma_s^2$ are the mean and variance of the shadowed LOS component, respectively.} This model has shown satisfactory agreement with  measured results in rural environments.

\paragraph*{Corazza-Vatalaro's Model \cite{Corazza1994}}
Unlike Loo's model, this statistical model combines Rician and log-normal distribution for the LOS signal, which is suitable for all environments. The model is tested for both LEO and MEO earth satellites, where the theoretical results match the measured results. In this model, the PDF of the signal envelope is the product of Rician and log-normal distributions, given by
\begin{equation}
f(r)  =  \int_{0}^\infty f(r|a) {  f_{\text{s}}(a)} \,\textrm{d}a,
\end{equation}
where
\begin{equation}
f(r|a) = \frac{r}{\tilde{K} a^2}\, \exp\left({-\frac{r^2}{2\,a^2 \tilde{K}}-K}\right)
I_0\left(\frac{r}{a} \sqrt{\frac{2K}{\tilde{K}}}\right),
\end{equation}
$K$ is the Rice factor,  $\tilde{K} \triangleq  0.5(K+1)$, {   and ${f_{\text{s}}(a)}$ is given in \eqref{eq: logpdf}}. Based on the values of $K$, the above model can be reduced to any non-selective fading models. This model was extended in \cite{Ko1997} by including the effects of phase variations in the shadowing and fading environment.


\paragraph*{Hwang's Model\cite{Hwang1997}}
This model extends the Corazza-Vatalaro's model by including the independent shadowing which affects both the direct and diffused LOS link components. The PDF of the signal envelope in this model is given by
\begin{equation}
f(r) = \int_{0}^\infty\int_{0}^\infty f(r|a_1,a_2) {  f_{\text{s}}(a_1) f_{\text{s}}(a_2)}\, \textrm{d}a_1\, \textrm{d}a_2,
\end{equation}
where
\begin{equation}
f(r|a_1,a_2) = \frac{r}{\sigma_m^2 a_2^2}\exp{\left(-\frac{r^2 + A^2 a_1^2}{2 \sigma_m^2 a_2^2}\right)} \,I_0\left(\frac{r A  a_1}{a_2^2 \sigma_m^2}\right)
\end{equation}
$a_1$, $a_2$ are the independent log-normal distributions for the direct and diffused LOS links, respectively, $A$  is the LOS component of fading. Note that $f(a_1)$ and  $f(a_2)$
are represented by \eqref{eq: logpdf} with parameters $\mu_{s1}$, $\mu_{s2}$, $\sigma_{s1}^2$, and $\sigma_{s2}^2$.
When $\sigma_{s1}^2 \rightarrow 0$, $\sigma_{s2}^2 \rightarrow 0$, and $F \rightarrow \infty$,  the fading component is absent, while for $\sigma_{s1}^2 = \sigma_{s2}^2$ and $\mu_{s1} = \mu_{s2}$, this model tends to follow Corazza-Vatalaro's model.
\begin{table*}
\caption{List of statistical models for LMS channel}\label{table:lms}
\centering
\begin{tabular}{|p{1.2cm}|p{1.0cm}|p{2.3cm}|p{1.7cm}|p{1.5cm}|p{7.2cm}| }
	\hline
	\hline
	\textbf{Ref.} &\textbf{Year}& \textbf{Multipath fading} &\textbf{Shadow fading}& \textbf{State}   &\textbf{Comments }\\
	\hline\hline
	\cite{Loo1985}&1985       &Rayleigh & Log-normal&Single & Applicable only for rural environment; does not consider the Doppler effect usually present in LEO satellites\\ \hline
	\cite{Corazza1994}&1994    &Rice \& Log-normal        &Log-normal    &Single& Applicable to both urban and rural environments because of the additional degree of freedom in modeling the LOS signal\\ \hline
	\cite{Saunders1996}        &1996   & - &Rayleigh&Single & Calculates the deep-fading probability in urban environment by incorporating height of the surrounding buildings, width of the street, and satellite geometry\\ \hline
	\cite{Hwang1997}           &1997  &Rice \& log-normal with independent shadowing       &Log-normal    &Single& Considers the log-normal shadowing independent from the multipath fading, allowing greater flexibility in fitting the real channel measurements compared to the Corrazz-Vatalaro's model \\ \hline
	\cite{Patzold1998}        &1998  &Rice        &Log-normal    &Single& Accounts for the Doppler effect,  making it a better candidate for LEO satellites including CubeSats  \\ \hline
	\cite{Fontan2001}         &2001  &Rayleigh &Log-normal    &Three& Elucidates a Markov chain-based geometrical model, showing a good
	agreement to the real measurements over several frequency bands (observed in all cases).\\ \hline
	\cite{Abdi2003}          &2003  &Rice &Nakagami    &Single& Provides a mathematically tractable model while conforming to the real measurements for both narrow-band and wide-band systems\\ \hline
	\cite{Scalise2006}          &2006  &Adaptive &Adaptive    &Multi-state& Describes a blind model, where the number of the Markov state and the distribution of the signal are not required apriori\\ \hline
	\cite{Kourogiorgas2014}          &2014  &Rayleigh &Inverse Gaussian    &Single& Experimentally investigates the effect of tree shadowing and introduces the inverse Gaussian distribution to better model the shadowing  \\ \hline
	\cite{Salamanca2019}          &2019  &Rayleigh (Adaptive) &Nakagami (Adaptive)   &Multi-state& Develops a finite Markov chain-based model, which adapts to the geometry of the CubeSats  \\
	\hline
	\hline
\end{tabular}
\end{table*}
\paragraph*{Patzold's Model \cite{Patzold1998}}
{  This model is similar to the Loo's model; however, it also considers the  Doppler frequency shift due to the relative motion between the Cubesat and  ground station. The Doppler shift can be approximated as
\begin{equation}
f_{\text D}= f_{\text c} \frac{v \cos(\phi)}{c}
\end{equation}
where $f_{\text c}$ is the carrier frequency, $\phi$ is the elevation angle, $v$ is the tangential speed of the satellite, and $c$ is the speed of light \cite{Popescu2016}. More accurate representation of the frequency shift can be found in \cite[(5)]{AliDhaHer:98}.
The Patzold's statistical model has an increased flexibility and fits well the measurements due to the realistic assumptions on the Doppler effect and the fading.

The PDF of the signal envelope is given by
\begin{equation}
f (r) = \int_0^\infty
\frac{r}{\phi_0}\exp\left(-\frac{r^2+a^2}{2\phi_0}\right) I_0\left(\frac{ra}{\phi_0}\right) f_{\text{s}}(a) \,\textrm{d}a,
\end{equation}
where $\phi_0 = \frac{2}{\pi} \sigma_0^2\,  \arcsin(\kappa_0)$, $\sigma_0$ is the mean power of the random process, and ${0<\kappa_0<1}$ is a parameter for the Doppler power spectral density, which can also control the fading rate. More precisely, the Doppler power spectrum density for the noise can be represented as
\begin{equation}
S(f) = \begin{dcases}
\displaystyle \frac{\sigma_{0}}{\pi f_{\text{Dmax}} \sqrt{1-(f/f_{\text{D}\max})^{2}}} & \text{for} |f|\leq \kappa_0 \\
0 &\text{for }|f|> \kappa_0
\end{dcases},
\end{equation}
where the parameter $\kappa_0$ determines the truncation frequency for function.}
The derived PDF in this model is a generalization of Rice density and therefore is more flexible. Also, this model is similar to the Loo's model; however, both models have different high order statistical properties, i.e., level crossing rates and average duration of fades.

\paragraph*{Kourogiorgas' Model \cite{Kourogiorgas2014}}
This model investigates the first order statistics for the LMS channel in two different tree shadowing scenarios, i.e., intermediate and heavy tree shadowing, respectively. Small unmanned aerial vehicle (UAV) was used as a pseudo-satellite to experimentally investigate the effect of the tree shadowing. It was shown experimentally in \cite{Kourogiorgas2014} that Loo's model offers the best accuracy among other models for the first order statistics of the received signal envelope. Furthermore, the authors also introduced inverse Gaussian (IG) distribution to model the tree shadowing \cite{Kourogiorgas2016}. Experimental tests were performed at a park, where the LOS signal was modeled as an IG distribution, while the multipath was modeled with a Rayleigh distribution. The PDF of signal envelope in \cite{Kourogiorgas2016} is given by
\begin{equation}
f(r) = \frac{r}{\sigma_m^2}\int_0^\infty\exp\left(-\frac{r^2+a^2}{2\sigma_m^2}\right)I_0\left(\frac{ra}{\sigma_m^2}\right)\tilde{f}_{\text{s}}(a)\, \textrm{d}a,\,\,r\geq 0
\end{equation}
where $\tilde{f}(a)$ is the PDF of the inverse Gaussian distribution given by
\begin{equation}
\tilde{f}_{\text{s}}(a) =\sqrt{ \frac{\lambda}{2\pi}} a^{-\frac{3}{2}}\exp\left(-\lambda\frac{ (a-\mu)^2}{2\mu_{s}^2 a}\right), ~~a>0.
\end{equation}
Here, $\lambda$ and $\mu_{s}$ are the parameters of the IG distribution with variance $\sigma_{s}^{2}=\mu_{s}^{3}/\lambda$.

\paragraph*{Abdi's Model \cite{Abdi2003}}
This model characterizes the amplitude of the LOS signal by Nakagami distribution. This model is more flexible due to the closed-form expressions of the channel statistics. The expression for the signal envelope in  \cite{Abdi2003} is given by
\begin{eqnarray}
f(r) &=& \left(\frac{\sigma^{2}_{m}\, m}{ \sigma^{2}_{m}\, m + \Omega}\right)^m \frac{r}{\sigma^{2}_{m}}\exp\left(-\frac{r^2}{\sigma^{2}_{m}}\right)\nonumber \\ && \times \phantom{}_{1}F_{1}\left(m, 1, \frac{\Omega\, r^2}{\sigma^{2}_{m}(\sigma^{2}_{m}+\Omega)}\right), ~~r\geq 0
\end{eqnarray}
where  $m \geq 0$ is the Nakagami parameter, $\Omega > 0 $ is the spread parameter, and $\phantom{}_{1}F_{1}(., ., .)$ is the confluent hyper-geometric function \cite{abramowitz1988}. This model fits well the Loo's model and the measured results, for both narrow-band and wide-band systems. It has an additional advantage over previous models by having closed-form expressions for the channel statistics, e.g., the PDF, CDF, and moment generating function; therefore leading to a more tractable analysis.

\paragraph*{Saunders' Model \cite{Saunders1996}}

A geometrical approach is used to determine the blockage probability of the direct path. The geometry of the streets and buildings,  which introduce a shadowing effect to the direct path, is taken into account. The probability that the direct path is blocked, i.e., when the height of the blocking building is larger than a certain threshold $h_t$, can be written as
\begin{equation}
P_b = \exp\left(-\frac{h_t^2}{2 \sigma_b^2}\right),
\end{equation}
where $\sigma_b^2$ is the variance of the building heights. By simple trigonometric relations, $h_t$ can be written as
\begin{equation}
h_t = \begin{dcases}
h_m + \displaystyle \frac{d_m \tan \phi}{\sin \alpha} & \text{for \hphantom{-}}~~0<\alpha\leq\pi \\
h_m + \displaystyle\frac{(w-d_m) \tan \phi}{\sin \alpha}&\text{for }-\pi < \alpha \leq 0
\end{dcases},
\end{equation}
where $h_m$ is the height of the receiver from the ground, $d_m$ is the distance between the building face and the receiver, $w$ is the width of the street, $\alpha$ is the azimuth angle between the receiver and the satellite, and $\phi$ is the elevation angle.
\subsubsection{Dynamic Models}
The dynamic models are based on Markov chains, with different states for the LMS channel, where each state corresponds to a different propagation environments.

\paragraph*{Fontan's Model \cite{Fontan2001}}
This model considers a three-state Markov chain for the three main propagation channel elements, i.e., the direct LOS, diffused LOS, and multipath signals.  Markov chain states are defined based on the degree of shadowing. This model is also tested for both narrow-band and wide-band conditions in which the multipath delays are assumed to be exponentially distributed. Fontan \textit{et al.} tested this model for L-band, S-band, and Ka-band frequencies in various environments with different elevation angles. They also developed a simulator that can generate a time series of channel parameters, including Doppler spectra, phase variations, power delay profiles, and signal envelopes.

\paragraph*{Scalise's Model \cite{Scalise2006}}
This model is based on the reversible-jump Monte Carlo Markov chain (RJ-MCMC) to characterize the  LMS channel. First-class statistical models work well under stationary conditions; however, they are un-satisfactory when substantial changes occur to the propagation channel. Also, the multi-state Markov chain-based models may not well-characterize the actual LMS channel. For example, in \cite{Fontan2001}, different states of the Markov chain represent the different channel elements, each having a fixed PDF. These assumptions make the model sensitive to the changes in the propagation environment. Hence, the RJ-MCMC model does not make any apriori assumptions about the propagation environment, the number of Markov states, or the distribution of the envelope. This model was tested at the Ku-band, where the model fit well with the measured results.

\paragraph*{Nikolaidis' Model \cite{Nikolaidis2017}}
This model uses a dual-polarized multiple-input multiple-output (MIMO) for measurement of the LMS channel, achieving the capacities between $4.1$-$6.1$ bits/second/Hz for both LOS and NLOS conditions. The channel capacity varies significantly with the received signal pattern and elevation angle. This stochastic channel model also approximates the channel capacities and correlation statistics. Moreover, a mean quasi-stationary time of $41$-$66$ seconds was found for different environments.

\paragraph*{Salamanca's Model \cite{Salamanca2019}}
A finite-state Markov channel with two sectors  was introduced in \cite{Salamanca2019} for LMS channel modeling. This adaptive model depends on the elevation angle of the satellite. More precisely, for low elevation angles, where the LOS signal is blocked, the fading amplitude is modeled by a Rayleigh distribution. On the other hand, a Nakagami PDF can describe the distribution of the LOS signal envelope at higher elevation angles. The performance of the communication system over the proposed channel was simulated in terms of the \ac{BER} and throughput, following the \ac{CCSDS} recommendations.

{  
\subsection{CubeSat-to-CubeSat Communications}
CubeSats can provide extended coverage in space and on Earth by working as inter-satellite relays. However, coordination among CubeSats requires C2C communications, which is a challenging task. The existing C2C link uses RF communications, highly-directed lasers, and VLC. The latter two require accurate pointing among the CubeSats,  while the first is not suitable for high-data-rate applications and systems with sensitive onboard electronics. Most of the missions that employ C2C communications are based on either RF or lasers.

The use of RF in high-frequency bands (e.g., SHF and EHF bands) can provide a solution for high-data-rate inter-satellite links. The main advantage of RF links is that they do not necessitate precise antenna-pointing mechanisms, as opposed to wireless optical links. However, RF technology requires high transmission power to compensate for the increased path loss at such high frequencies. Also, the RF technology is subject to performance degradation due to the interference between neighboring C2C links operating on the same frequency band.  The C2C RF-based link is considered to be a LOS communication. Unlike the C2G channels that experience fading, the power of the received signal can be regarded as fixed, as it depends mainly on the free space-path loss.  The most ambitious mission that uses RF-based C2C links was QB-50, which consisted of a swarm of fifty CubeSats for the purpose of studying the Earth's upper thermosphere. Until that point, a swarm of $36$ CubeSats was launched, utilizing RF-based C2C communications \cite{Bedon2010}. The feasibility of C2C links is numerically evaluated in \cite{Danil2017} where it is shown that using higher frequencies reduces the communication time for C2C links. An inter-satellite scheme that employs channel-coded CDMA has been proposed in \cite{BauCas:13}.  The antennas  have been designed to maintain a fixed link irrespective of their orientations on the CubeSat faces, offering a $5$~dBi gain.
 Recently, in \cite{Bulanov2018}, the authors investigated the QoS requirements for the C2C link with massive MIMO. The results in \cite{Bulanov2018} showed that using massive MIMO for C2C links improves communication time; however, the use of massive MIMO increases the size of the CubeSats and the power consumption. A swarm of four CubeSats was developed in \cite{Yoon2014}, which demonstrates multipoint to multipoint high data rates C2C links.  S-band frequencies were used in \cite{Yoon2014} achieving 1 Mbps downlink and 100 kbps crosslink data rates.

On the other hand, lasers is a promising technology for establishing C2C links, as they allow high-data-rate communication. In \cite{SmuKate:09}, Smutny \textit{et al.} tested the optical link between two LEO satellites. These links allow ultra-reliable communication with a bit error rate of less than $10^{-9}$, while operating at a data rate of up to $5.6$ Gbps.
Unlike laser C2G links, the optical beam for C2C communication in free space channels is not subject to atmospheric turbulence. However,   the main issues that negatively impact the performance of laser-based C2C links are
\begin{itemize}
	\item Pointing loss from the imperfect acquisition and tracking, since the two satellites move with different relative velocities.
	\item Doppler frequency shift from the relative motion between the CubeSats.
	\item Background noise from the Sun and other stars, and receiver noise.
\end{itemize}
The direction of the laser beam can be corrected with the help of beam-steering mirrors. Various tracking techniques such as DC tracking, pulse tracking, square law tracking, coherent tracking, tone tracking, feed-forward tracking, and gimbal tracking can be used for inter-satellite laser links.

Besides RF and lasers-based C2C links, recent advancements in free-space optics and light-emitting-diodes (LED) technologies have triggered the use of VLC for C2C links. LED technology is advantageous over its counterparts  becasue of its low power consumption and lightweight. In \cite{Wood2010}, Wood \textit{et al.} examined the feasibility of using LEDs for a hypothetical C2C link, focusing on minimizing the background illumination for these links.
In addition to the background illumination noise, in \cite{Amanor2018} and \cite{Amanor2017}, Amanor \textit{et al.} investigated the effect of solar radiation on VLC-based C2C links and found that the solar radiation significantly reduced the SNR of the received signal. Using a transmit power of $4$ Watts and digital pulse-interval modulation, a data rate of $2$ Mbps was achieved with BER = $10^{-6}$ for a transmission range of $500$ meters. The proposed scheme was designed to comply with the limited size, mass, power, and cost requirements of the CubeSats.

}
{  
\subsection{Lessons Learned}
In this section, we presented the two types of CubeSat communication links, i.e., C2G and C2C, and their corresponding channel models. Two leading competent-communication technologies can be used to establish the links: optical and RF.
For C2G, laser communication offers a high data rate; however, it suffers from pointing error and atmospheric turbulence. Therefore, RF systems are currently preferable over optical communication for C2G links.

For RF C2G links, we presented various channel models that can be used in the context of CubeSats. These models fall into two main categories: static and dynamic frameworks.  In the static models, the signal envelope can be modeled by a single distribution that does not change with time. Hence, it makes sense to describe static or stationary channels. On the other hand, dynamic models are represented by a mix of several statistical distributions. Dynamic multi-state models are more appropriate for the CubeSats than static single-state models because of the continuous movement of the satellite and the coverage of large areas that experience distinct shadowing and multipath effects.  Hence,  finite Markov chain-based models that are adaptable to the geometry of  CubeSats can be considered as a competent-candidate for the channel model.  Also, considering that the LOS components undergo Nakagami fading permits us to represent the signal statistics in a closed-form, which in turn facilitates  theoretical analysis.  Moreover, accounting for the Doppler effect leads to more realistic channel models.

With respect to the C2C links, optical communication offers a promising solution, in which the radiated beam is not subject to atmospheric turbulence. However, efficient tracking, acquisition, and stabilizing mechanisms are required to ensure the reliability of the link. Low-cost light-pointing systems should be developed to cope with the limitations on the size, mass, and power of CubeSats.
}
\section{Link Budget}
Generally speaking, establishing a reliable communication link between a transmitter and a receiver is the ultimate goal of radio-link design. In particular, a CubeSat establishes two types of duplex radio links, uplink and downlink, with ground stations and with other CubeSats. Despite the key role of the communication subsystem, the power that a CubeSat can dedicate is limited by its weight and size constraints \cite{Benson2017}.  This section discusses the link budget expression for the downlink, i.e., CubeSat-to-ground communications.  The link design must ensure the ability to transmit and receive data directly from space to Earth or through one or more communication relays \cite{Latachi2017,Alvarez2016}.
A link budget is a set of parameters that define a communication link in terms of the power available for a reliable connection between the transmitter and  receiver. {  The satellite-to-ground link (downlink)'s energy-per-bit to noise spectral density, which measures the reliability of the link, is calculated based on the link budget. The energy-per-bit to noise spectral density at the ground station can be expressed as
	\begin{equation}\label{eq:LinkBudget}
	\frac{E_{b}}{N_{o}} = \frac {P_{\text t}\, G_{\text t}\, G_{\text r}}{L\, k T \,R_{\text b}},
	\end{equation}
	where $P_{\text t}$ is the transmitted power, $G_{\text t}$ and $G_{\text r}$ are the transmitter and receiver antenna gains, $T$ is the system temperature noise, $R_{\text b}$ is the target data rate, $k$ is the Boltzmann constant, and $L$ is the overall loss. The overall loss accounts for the losses occurred while the signal propagates from the satellite to ground station, which can be attributed to four main components as follows:
	\begin{itemize}
		\item Free-space path loss, $L_{\text p}$, because of the  basic power loss that increases inversely with the square of the distance propagated.
		\item Atmospheric loss, $L_{\text{a}}$,  due to absorption and scattering of the field by atmospheric particulates, for instance, signal  attenuation caused by rainfall.
		\item Polarization loss, $ L_{\text{pol}} $, due to an improper alignment of the receiving antenna subsystem with the received wave polarization, leading to polarization mismatch.
		\item Antenna misalignment loss, $L_{\text{aml}}$, due to the difficulty of steering to the ground station antenna in exactly the correct direction of the CubeSat. 
	\end{itemize}
	More precisely, the overall loss $L$ can be represented as
	\begin{equation}
	L=L_{\text p} \, L_{\text{a}}\, L_{\text{pol}}\, L_{\text{aml}}.
	\end{equation}
}
\begin{figure}
	\centering
	\includegraphics[width=0.8\columnwidth]{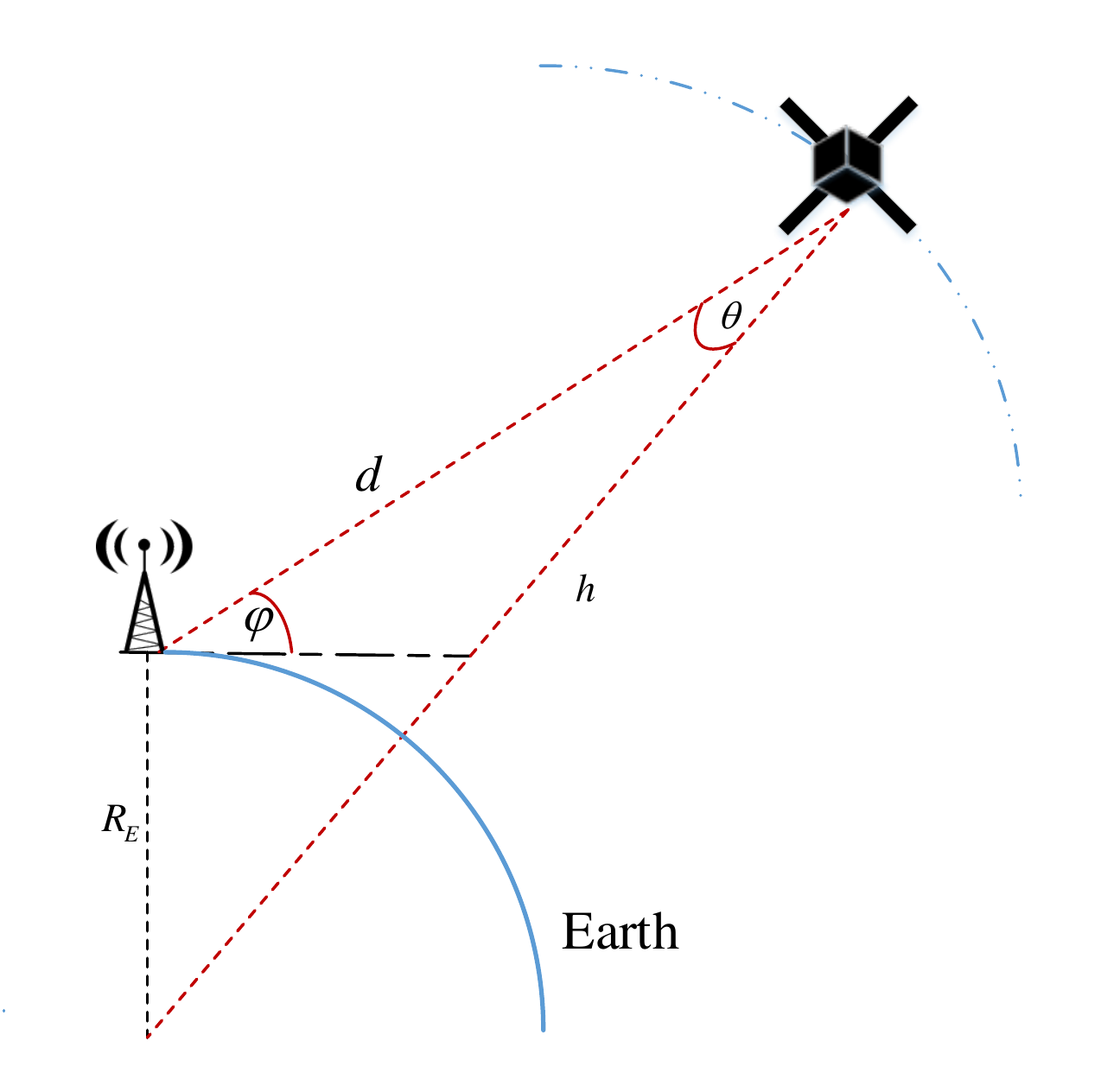}
	\caption{Schematic description of a LEO CubeSat trajectory}\label{fig: traj}
\end{figure}
%
The free-space path loss $L_p$ is given by
\begin{equation}\label{eq: fspl}
L_p =  {\left(\frac{4 \pi d}{\lambda}\right)}^{2},
\end{equation}
where $d$ is the distance between the ground station and the satellite and $\lambda$ is the wavelength of the signal. Note that $d$ depends on the parameters of the LEO orbit such as the minimum elevation angle $\phi$, angle between the position of CubeSat in orbit and the ground station $\theta$, and CubeSat's altitude $h$ from the center of Earth. Based on these parameters, $d$ is calculated as \cite{ALMONACID201795}
\begin{equation}
d = \sqrt{(R_E + h)^2 -R_E^2\cos^2\phi}-R_E \sin\phi,
\end{equation}
where $R_E$ is the Earth radius \cite{Popescu2016}. Fig. \ref{fig: traj} depicts the relationship between these parameters and the distance.
\begin{figure}
	\centering
	\includegraphics[width=0.85\linewidth]{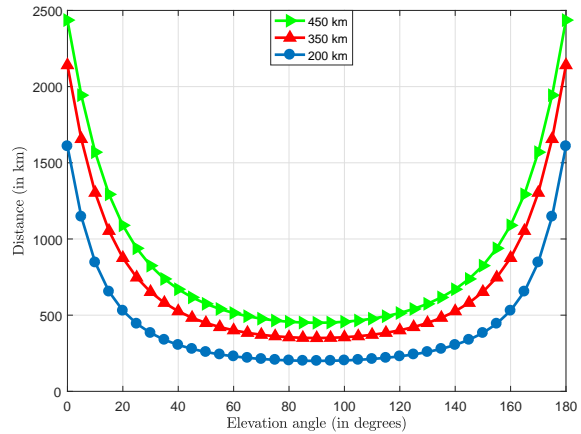}
	\caption{Impact of the elevation angle on the distance between the satellite and the ground station.}\label{fig: sim1}
\end{figure}
For illustration purposes, we consider LEO orbits with three different altitudes and calculate the distance between the satellite and the ground station as shown in Fig. \ref{fig: sim1}. It is clear from Fig. \ref{fig: sim1} that the distance between the ground station and the satellite is minimal when the elevation angle is $90$ degrees. To best characterize the effect of the elevation angle on the path loss, we consider VHF-band and L-band frequencies and calculate the path loss with respect to the elevation angle as shown in Fig. \ref{fig: sim2}. It is clear that the path loss is low at the $90$-degree elevation angle due to the shorter distance. Also, at higher frequencies, the path loss increases with the altitude of the satellite.
\begin{figure}[h]
	\centering
	\includegraphics[width=0.89\columnwidth]{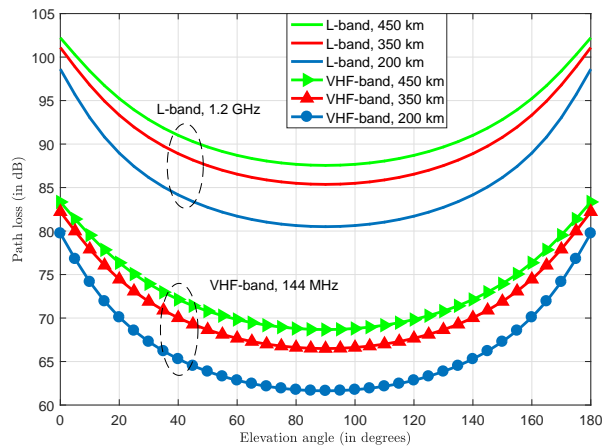}
	\caption{Impact of the elevation angle on the path loss at different frequency bands.}\label{fig: sim2}
\end{figure}
{  
	On the other hand, the atmospheric attenuation due to water vapor absorption and rain attenuation  for various frequency bands can be predicted as in \cite{DisAllHai:97,Atmospheric:12}. Polarization loss occurs when a receiver is not matched to the polarization of an incident electromagnetic field. For linear polarization, the  loss is expressed as a function of  the angle $\Theta$ between the  polarization vectors at the transmitting and receiving antennas\footnote{For other types of polarization, the interested reader can refer to \cite{Polarization:96,GalRasHaz:06}.}.
 The antenna misalignment loss, $L_{\text{aml}}$ can be represented as a function of the pointing-error angle, $\phi_{e}$ and the antenna beamwidth $\phi_{b}$. For example, in the case of a parabolic antenna, the  antenna-misalignment loss  can be calculated as
	\begin{equation}
	L_{\text{aml}}= \exp\left(2.76\, \frac{\phi_{e}^{2}}{\phi_{b}^{2}}    \right).
	\end{equation}
	It is clear that the misalignment loss is inversely proportional to the antenna beamwidth\cite{Gar:12}. In RF-based communications, the antenna misalignment loss is usually less than $1$~dB; therefore, it can be included in the link margin.  However, for laser communication between the CubeSat and  ground station or for inter-CubeSat links, a tiny pointing error can lead to severe degradation in performance, due to the narrow beamwidth. Therefore, accurate pointing techniques are essential for optical communication. The pointing error angle can be reduced by employing gimbals, permitting precise pointing between the CubeSat and ground-station antennas. NASA developed the rotary-tilting gimbal for directional control of CubeSats \cite{NASAGimbal:17}.
	
In Fig.~\ref{fig:snr}, the energy-per-bit to noise spectral density at the ground station is shown versus the CubeSat altitude for various frequency bands and elevation angles. The parameters of the link budget are specified in Table~\ref{linkbudgetparam}.
	\begin{table}[]
		\caption{  Parameters of the link budget calculation}
		\centering
		{  	\begin{tabular}{|l|l|}
				\hline
				{\bf	Parameter} & {\bf Values} \\ \hline
				Transmitted power $P_{t}$		&  $15$ dBm             \\
				CubeSat antenna gain $G_{t}$ 	&  $0$ dBi      \\
				Ground station antenna gain $G_{r}$	&   $12$ dBi     \\
				Total noise temperature $T$	&     $1160$ K   \\
				Data rate $R_{\text b}$	& $2.4$ kbps       \\
				Beamwidth $\phi_{b}$	& $2.9^{\circ}$       \\
				Pointing angle error $\phi_{e}$	&  $0.5^{\circ}$      \\
				Pointing loss $L_{\text{aml}}$	&  $0.35$~dB      \\
				Polarization mismatch loss $L_{\text{pol}}$	&  $1$~dB      \\
				Atmospheric loss $L_{\text{a}}$	&  $2.5$ ~dB      \\ \hline
		\end{tabular}}
		\label{linkbudgetparam}
	\end{table}
From Fig.~\ref{fig:snr}, we can see that the signal quality also depends heavily on the elevation angle.
}

The generalized SNR expression in \eqref{eq:LinkBudget} is valid for most of CubeSat missions. However, the path loss varies for different missions due to the geographical location of ground stations, the operating frequency band, varying attenuation, and the altitude of the orbits \cite{Vertat2012}. The frequency allocation for the CubeSat links, i.e., space-Earth, Earth-space, and inter-satellite, is regulated for different applications by international entities. Typically, the frequency bands that are used for CubeSats are very-high frequencies (VHF) or ultra-high frequencies (UHF) amateur bands \cite{Barbaric2018}. However, some of the missions also use Ka-band \cite{kegege}, X-band \cite{peragin2016, Palo2015}, S-band \cite{Ceylan2011}, L-band \cite{Barbaric2018} and optical waves.  Further, in \cite{AKYILDIZ2019166}, Akyildiz \textit{et al.} proposed a multi-band radio which covers a wide range of spectra including microwaves, mm-waves, THz band, and optical waves for CubeSats. The link budget was calculated to show the effectiveness of the multi-band radios with continuous global coverage for IoT networks.
Table \ref{table:satfreq} summarizes the different frequency bands used for CubeSats \cite{kulu_2019}.
\begin{figure}
	\centering
	\includegraphics[width=0.92\linewidth]{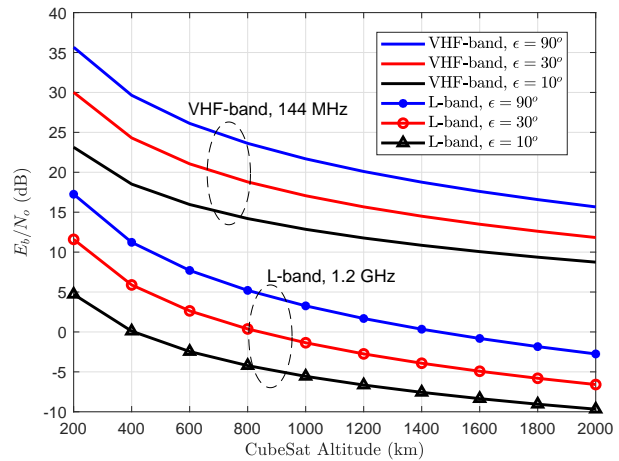}
	\caption{The energy-per-bit to noise spectral density for the downlink at the ground station vs the CubeSat altitude for various frequency bands and elevation angles.}
	\label{fig:snr}
\end{figure}
\begin{table}
	\caption{Frequency bands of CubeSat missions}\label{table:satfreq}
	\centering
	\begin{tabular}{|p {2.3 cm}|p{5.3 cm}|}
		\hline
		Frequency band      &Frequency range\\
		\hline
		HF                      &20-30 MHz\\
		VHF                      &145-148 MHz\\
		UHF                      &400-402, 425, 435-438, 450-468, 900-915, and 980 MHz\\
		L-Band                   &1-2 GHz\\
		S-Band                   &2.2-3.4 GHz\\
		C-Band                   &5.8 GHz\\
		X-Band                   &8.2-10.5 GHz\\
		Ku-Band                  &13-14 GHz\\
		K-Band                   &18 GHz\\
		Ka-Band                  &26.6 GHz\\
		W-Band                   &75 GHz\\
		Optical            &400-700 THz \\
		
		\hline
	\end{tabular}
\end{table}

\section{Modulation and Coding}
A fundamental feature of CubeSat communication systems is the design of the modulation and coding schemes. Since the weight and  cost of CubeSats are limited, there are major constraints on the transmitted power. Hence, achieving a reliable communication with limited energy over LMS channels is a challenging issue. Depending on the CubeSat mission, the design of the modulation and coding schemes should take into account the proper trade-off between several parameters. These parameters can be summarized as follows:
\begin{itemize}
	\item   The operational frequency band, e.g., the UHF, S, X, and Ka bands, and the  allocated bandwidth.
	\item   The target data rate.
	\item    The duration of ground passes (i.e., the period during which the CubeSat is able to communicate with the ground station).
\end{itemize}

For example, the available bandwidth at a higher frequency band such as the X-Band can reach $375$~MHz, while the target bit rate is on the order of $150$~Mbps for typical earth-exploration CubeSat missions. Hence, binary modulation methods,  along with low-rate channel codes  with high error-correction capabilities, are preferable over higher-order modulation schemes with high-rate \ac{FEC} codes. This is attributed to the reduction in the required power in the former case with the existence of more redundant data for efficient error correction, leading to higher power efficiency.

On the contrary, the available bandwidth at S-band for NASA missions is $5$~MHz. Therefore, higher-order modulations, e.g.,  $8$-\ac{PSK} with rate-$7/8$ LDPC code, are essential to improve spectrum efficiency \cite{Schaire2016,wong2016optimum}.

Other important features are the data volume needed to be communicated and the duration of ground passes. In fact, for some missions, the pass period is short while the amount of generated data is large. Hence, bandwidth-efficient communication systems with high data rates are required for reliable delivery of the information; thereby reducing the number of passes required. {  For example, the CubeSat for precipitation monitoring in \cite{RainCube:18} generates a daily payload of $1.73$ Gb, while the available data rate is $50$ kbps. Also, the spacecraft has a transmission duty cycle of only $25\%$, in order to be compliant with the power limitation. At an altitude of $450$ km, the CubeSat has a pass period of $10.8$ min, requiring $53$ passes to deliver its payload data. This mission employs a distributed network of $25$ ground stations. In order to reduce the number of required stations, the  data rate should be increased by employing a higher spectrum-efficiency modulation.}

Generally, choosing a suitable CubeSat modulation technique requires a trade-off between several metrics, i.e., the bandwidth and power efficiency, the \ac{BER} performance, and the complexity of the spacecraft transceiver. In the following, an overview of the most common CubeSat modulation and coding schemes is presented.

Several modulation schemes  such as \ac{QPSK}, \ac{OQPSK}, $M$-\ac{PSK}, $M$-\ac{APSK}, $M$-\ac{QAM} with $M\in \{4,8,16,32\}$, are used in CubeSats. The performance of these schemes was investigated in  \cite{Schaire2016}  for various \ac{FEC} coding rates and channel impairments like non-linearity. It was shown that higher-order modulations are vulnerable to non-linear distortion resulting from the CubeSat's power amplifier, with the exception to $16$-\ac{APSK}, as this latter requires only a quasi-linear power amplifier.


The authors in \cite{Munir2016,KirFau:18} suggest the use of \ac{GMSK}, where the CubeSat's payload is designed to cope with the link budget constraints for the system. It was shown through simulations that the CubeSat receiver can demodulate signals with received power as low as $-102.07$~dBm \cite{Munir2016}. This is attributed to the fact that \ac{GMSK} signals have  better spectral characteristics than \ac{OQPSK}. Also, they have constant envelopes, allowing amplifiers to operate in the saturation region, which increases their power efficiency; however, \ac{GMSK} has poor error performance when compared to \ac{OQPSK}\cite{Gaysin2017}. In order to improve the \ac{BER}, the authors in \cite{Gaysin2017} proposed  employing a Viterbi decoder, leading to higher computational complexity at the receiver.

\ac{OQPSK} and rotated $\pi/4$-\ac{QPSK} were proposed as possible modulation techniques for CubeSats in \cite{Gaysin2017} due to their robustness to the non-linearities of the amplifier, leading to better \ac{BER} performance than classical \ac{QPSK}. To improve the spectral characteristics of the \ac{OQPSK} (i.e., reducing  out-of-band emissions), the \ac{CCSDS} standard recommends using a filtered \ac{OQPSK} scheme implemented using a linear-phase modulator, i.e., \acsu{OQPSK/PM}.  The \ac{OQPSK/PM}-modulated signals have a constant envelope, permitting highly-efficient nonlinear power amplification \cite{RFCCSDS:18}.

The effect of the Doppler frequency shift on \ac{OQPSK/PM} was investigated for a constellation of CubeSats around the Moon in \cite{Babuscia2016}. More precisely, the maximum Doppler frequency and Doppler rate profiles were estimated, and, accordingly, a frequency-tracking loop was designed to track the Doppler frequency and rate.

For higher-spectrum efficiency, a hybrid modulation scheme can be used, where two parameters of the carrier, i.e., the frequency and  phase are simultaneously modulated. For instance, in \cite{Vertat2013}, Vertat \textit{et al.} proposed a hybrid $M$-\ac{FSK}/Differential-\ac{QPSK} modulation technique, leading to higher spectrum efficiency compared to  $M$-\ac{FSK}. Another means of achieving highly-efficient CubeSat communication systems is through the joint design of higher-order modulation schemes with error-correcting codes,  usually referred to as coded modulation frameworks, for example, \ac{TCM} \cite{Ungerboeck:87}. One of these techniques is the bi-dimensional $M$-PSK-\ac{TCM}, which depends  on $M$-\ac{PSK} modulations with the use of \ac{CC} to introduce legitimate sequences between signal points joined by the trellis of the code.\footnote{Note that bi-dimensional refers to the in-phase and quadrature components ($I/Q$).} A generalization of this scheme involves several parallel modulating sequences (more than two),  referred to as $L$-dimensional $M$-PSK-TCM.  In this technique, the joint design of \ac{CC}, $M-$\ac{PSK}, and multidimensional signal spaces provides a significant power gain, when compared to their sequential implementation, as shown in \cite[Fig.~B-4]{ModCodCCSDS:18}. A comparison between \ac{TCM}, \ac{CC}, and turbo coding is conducted with OFDM signaling for LEO satellite channels within the L-band and Ka-bands in \cite{FerRaj:98}. Turbo-coded-OFDM achieves the lowest \ac{BER} compared to \ac{CC}-OFDM and \ac{TCM}-OFDM systems.

CubeSat link performance can vary significantly during the communication window because of environmental conditions, for instance, rain or due to a change in the elevation angle of the observer ground station. For example, when the CubeSat rises from a $10^{\text{o}}$ elevation up to $90^{\text{o}}$, a variation in the $E_{b}/N_{o}$ up to $12.5$~dB can be seen, as shown in Fig.~\ref{fig:snr}. Hence, adaptive modulation and coding schemes are required, as they can offer efficient communication over a wide range of signal-to-noise ratios through the proper interplay of  power and spectrum efficiency. For adaptive modulation,  the received signal strength should be calculated correctly at the satellite. In  \cite{Vertat2012}, a carrier-to-noise ratio ($C/N_0$) estimator was proposed based on a fast Fourier transform.

To further increase the link performance, traditional \ac{MIMO} schemes with multiple antennas in both the transmitter and  receiver are usually deployed in terrestrial systems. However, most CubeSats cannot support multiple antennas due to size and cost limitations. Also, the separation between the antennas of the CubeSats  would be too small to permit high-performance gains, in other words, the channel is rank-deficient.  Therefore, cooperative-communication techniques  can be employed for CubeSats, whereby different spatially-distributed spacecrafts, each with a single antenna, work together as a virtual entity. In this regard,  a space-time-based scheme, i.e., Alamouti's code, was proposed in \cite{Gibalina2018} for CubeSats to achieve high-diversity gain. The BER performance of distributed $2 \times 2$  \ac{MIMO} and $2\times1$ \ac{MISO} schemes was simulated for various channel codes (convolutional, Reed-Solomon, LDPC, and turbo codes). It was found that combining distributed \ac{MIMO} with channel coding leads to better error performance compared to \ac{SISO} based schemes. On the other hand, these techniques face several challenges, including phase synchronization between different satellites and induced latency in the inter-satellite links used for cooperation between CubeSats.

Given the complexity of the communication system, there are more constraints on the computational power of the transceivers in CubeSats compared to those in ground stations. These limitations derive from the finite power available for computations in the satellite due to its limited size and cost. In this regard, a  communication scheme between lunar CubeSat network consisting of $20$ spacecrafts, and an earth station was proposed in \cite{Babuscia2015}. The system aims to achieve multiple-access communication with a trade-off between the complexity of the CubeSat and that of the earth station. More precisely, the ground station uses un-coded CDMA  in the uplink, allowing for the possibility of having a simple decoder at the CubeSat. On the other hand, for the downlink, the spacecraft employs a low-complexity sparse LDPC encoder alongside a spread-spectrum transmitter, leading to higher power efficiency at the expense of increasing the complexity at the earth station.

\begin{table}
	\caption{The \ac{CCSDS} recommendations for modulation and coding schemes in LEO satellites.}
	\label{tab:modulation}
	\begin{tabular}{|p{1.5cm}|c|p{3.7cm}|}
		\hline
		Frequency Band	& Mission type  & Modulation techniques  \\
		\hline
		S-Band	& Space Research  &  \ac{GMSK}, filtered \ac{OQPSK}\\
		&  Earth Exploration  &  \ac{GMSK}, filtered \ac{OQPSK}\\
		\hline
		X-Band	& Space Research &\ac{GMSK}, filtered \ac{OQPSK}  \\
		
		& Earth Exploration  &$4$D $8$-\ac{PSK} TCM, \ac{GMSK}, filtered \ac{OQPSK}, $8$-\ac{PSK}, $M-$\ac{APSK} with $M\in \{16,32,64\}$   \\ \hline
		Ka-Band	& Space Research &\ \ac{GMSK} with precoding  \\
		
		& Earth Exploration  & \ac{GMSK}, filtered \ac{OQPSK}, $8$-\ac{PSK}, $M-$\ac{APSK} with $M\in \{16,32,64\}$ \\
		\hline
	\end{tabular}
\end{table}
Finally, we shed light on the \ac{CCSDS} recommendations for modulation and coding schemes to be employed in satellites  communication systems \cite{ModCodCCSDS:18}. Table~\ref{tab:modulation} shows the suggested modulation techniques for two types of missions, space research and earth exploration, operating at various frequency bands.

{  
\subsection{Lessons learned}
In this section, we reviewed various modulation and coding techniques employed in CubeSat systems. It is a challenging task to select the  proper  modulation and coding scheme to be employed in CubeSats, as this latter relies  on several key parameters, namely, the available transmission power, the allocated bandwidth, the target data rate, the generated payload data volume, and the duration of ground passes.  More precisely,  the  bandwidth allocated to missions operating at higher frequency bands (e.g., X- and Ku-bands) is quite large, allowing for a high  data rate, even with modest transmission power and binary modulation. Conversely, CubeSats operating at lower frequencies (e.g., S-Band) are assigned limited bandwidths, restricting the data rate. Consequently, if the generated data volume is very high, the spectrum efficiency should increase by using higher-order modulation schemes.

}
{
	  
	\section{Medium Access Control (MAC) Layer}
	MAC-layer protocols are vital for distributing resources efficiently throughout the network. Designing MAC-layer protocols for heterogeneous space-information networks is especially challenging because of users' varying demands, the diverse types of CubeSats and C2C links, and the uncertain space environment. The cooperative Earth-sensing model was introduced to replace the use of single satellites. Fig.~\ref{fig: SIN} illustrates a heterogeneous complex space-information network with conventional satellite networks, small satellites, and inter-networking between these networks. In such a model,  the MAC protocols play an important role in system performance. The protocol should take into account various parameters, such as the on-board power, the network topology, the mission objective, the available computing resources, and the number of satellites.
\begin{figure*}
\begin{center}
\includegraphics[width=1.5\columnwidth]{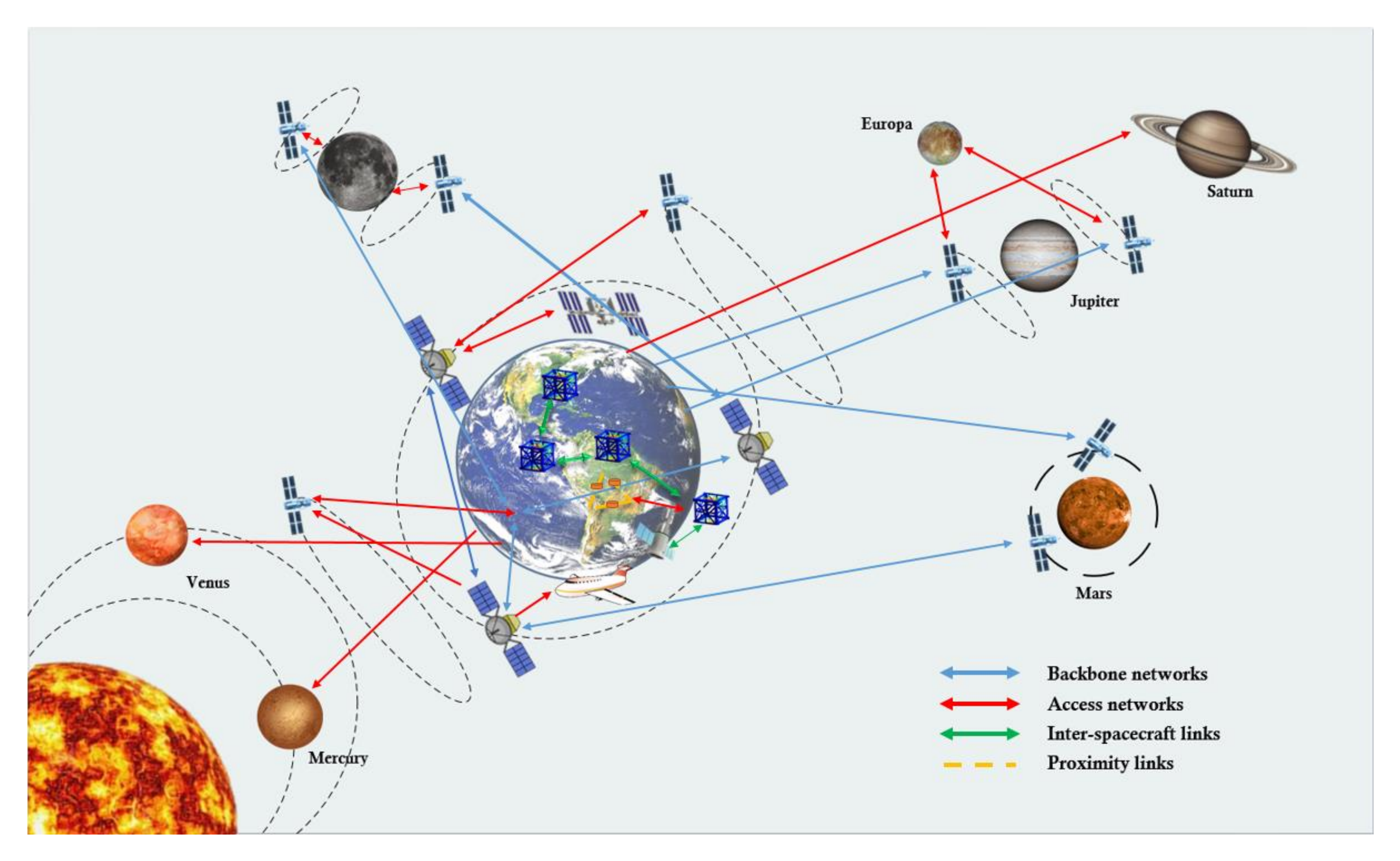}
 \caption{Heterogeneous space-information network.}
 \label{fig: SIN}
\end{center}
\end{figure*}	
	Recently, in \cite{Du2016}, Du \textit{et al.} discussed the architecture for complex heterogeneous space-information networks and provided protocols for sending the sensed data from the satellite to the ground in due time.   In \cite{radhakishnan2014}, Radhakishnan \textit{et al.} proposed a carrier-sense multiple-access with collision-avoidance, request-to-send, and clear-to-send (CSMA/CD/RTS/CTS) protocol  for CubeSats,   considering three different network setups, namely, cluster, leader-follower, and constellation. They performed various simulations for the $1$U CubeSats operating in the S-band with a transmission power of $500$ mW to $2$ W. Their results suggested that the average access delay and end-to-end delay were significant for a cluster network compared to the constellation and leader-follower setup. This was caused mainly by the conflict from  the use of the same frequency band by a greater number of CubeSats in close vicinity. Since this protocol delay is quite significant, it can only be used in missions that can tolerate delay.
	
	In \cite{radhakrishnan2014optimal}, Radhakishnan \textit{et al.} introduced a hybrid TDMA/CDMA protocol  for a swarm of CubeSats where each swarm has a master CubeSat and its respective slave CubeSats. The slave satellites only communicate with the master satellite within the swarm, whereas the master satellites forward the data to the next master satellite or to the destination. In the TDMA approach, the authors assign each cluster with a unique code through which each slave satellite dedicates time slots for downlink and uplink communication with the master satellite \cite{radhakrishnan2014optimal}. Alternatively, in the CDMA approach, satellites are assigned unique orthogonal codes that enable communication with the master satellite simultaneously, without interference. The hybrid TDMA/CDMA protocol can support large-scale satellite networks and has a low delay. In \cite{Pinto2015}, Pinto \textit{et al.} studied the direct sequence-code division multiple access (DS-CDMA) for inter-satellite communications. The performance of DS-CDMA was evaluated in terms of BER in the AWGN and Rayleigh environments, in which it was observed that increasing the number of users results in a low reliability. Recently, in \cite{Anzagira2018}, Anzagira \textit{et al.} introduced non-orthogonal multiple access  (NOMA) scheme  for VLC-based inter-satellite links. NOMA was able to achieve high reliability: a BER of $10^{-6}$ for a cluster of small satellites with a transmission power of $2$ W. Table \ref{table:maclayer} summarizes the MAC-layer protocols for CubeSats.
	\begin{table}
		\caption{{  MAC layer protocols for CubeSats}}\label{table:maclayer}
		\centering
		{  \begin{tabular}{|p{0.7cm}|p{1.8cm}|p{2.0cm}|p{2.7cm}| }
				\hline
				\hline
				\textbf{Ref.} &\textbf{Protocol}& \textbf{Network setup} &\textbf{Comments}\\
				\hline\hline
				\cite{Du2016}       &- &Cluster and constellation& Introduces heterogeneous space information networks \\ \hline
				\cite{radhakishnan2014}       & CSMA/CD/ RTS/CTS &Cluster, leader-follower, and constellation& High end-to-end delay   \\ \hline
				\cite{radhakrishnan2014optimal}       & Hybrid TDMA/CDMA &Cluster & High scalability and low end-to-end delay   \\ \hline
				\cite{Pinto2015}       & DS-CDMA &Cluster& High reliability for proximity links  \\ \hline
				\cite{Anzagira2018}      & NOMA &Cluster, leader-follower, and constellation& VLC-based inter-satellite links with high reliability  \\ \hline
				\hline
				\hline
		\end{tabular}}
	\end{table}	
}



\section{Networking}
{  
In CubeSats communications, networking is  vital for selecting the best path for reliable communication between the source and the destination node. Routing protocols based on various link performance metrics, such as bandwidth, reliability, latency, and transmission power facilitate determining the most appropriate paths for data delivery. Since CubeSats have low power capabilities, multi-hop routing can significantly reduce their power consumption.  Various routing protocols based on the type of CubeSat constellation have been presented in past literature. For instance, in \cite{Radhakrishnan2013}, Radhakishnan \textit{et al.} proposed the Bellman-Ford algorithm  for selecting the shortest path in a leader-follower CubeSat flying-formation pattern.
Similarly, Bergamo \textit{et al.} investigated several protocols for route discovery and synchronization in small satellites \cite{BERGAMO2005725}. They categorized possible neighbors into two types, new or old, depending on whether the satellite performing neighbor discovery has no frequency, velocity, and coordinates information about the new satellite or whether this information about the old satellite is already known. In \cite{Zhang2011}, Zhang \textit{et al.} developed a bandwidth-and-delay-aware routing protocol,  which constrains the delay when the bandwidth is overloaded, and uses another link when there is available residual bandwidth. Li \textit{et al.} proposed a routing protocol for small satellites that avoids using  invalid links and selecting only feasible links \cite{Yang2008}.  The feasibility of the links is estimated in advance by using an offline initialization strategy. They introduced the concept of a rectilinear Steiner tree for multicast routing in LEO satellites, which minimizes the total available bandwidth \cite{Yang2008}. This method consumes less bandwidth than the shortest path algorithm.
Di \textit{et al.} proposed a dynamic routing protocol that considers the satellite network as a mobile ad-hoc network \cite{Di2005}. This protocol divides the satellite network into clusters, assuming that the topology of the system is known. Once the network is divided into groups, an asynchronous transfer-mode-based routing protocol is used to enable virtual-path connections. This protocol has a low system overhead,  low latency, and high scalability.
\begin{figure}[h]
	\begin{center}
		\includegraphics[width=1\columnwidth]{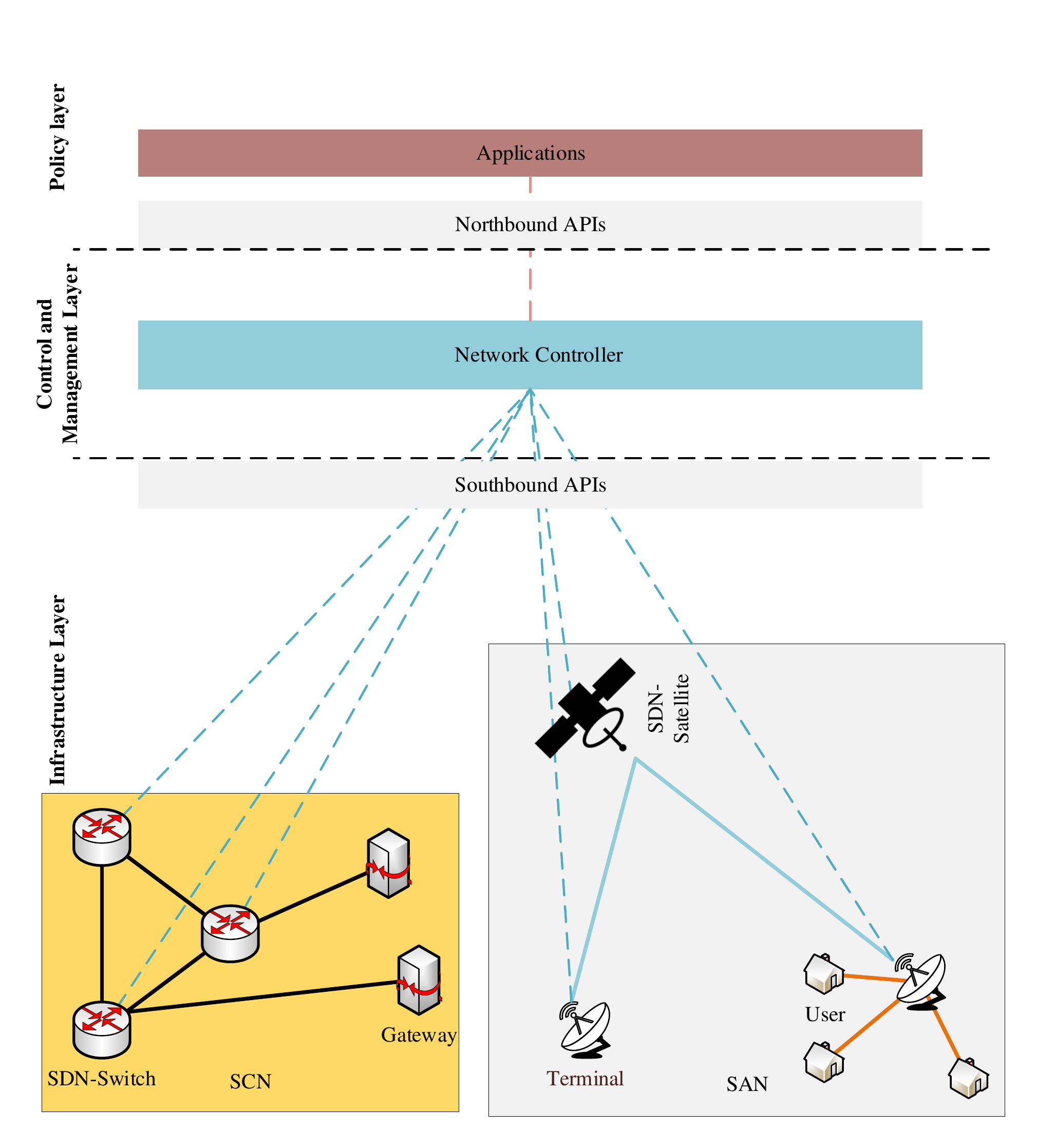}
		\caption{{  SDN-enabled architecture for small satellites.}}
		\label{fig: sdn-iost}
	\end{center}
	\vspace{-0.7em}
\end{figure}

Nevertheless, in practice, the satellites are not active all the time, which causes them to store their data for a long time,  leading to the development of delay-tolerant networking (DTN) paradigm. In DTN, the satellites store the data till the next available contact with the satellite or the ground. However, CubeSats' limited storage and capacity, prevent them from transmitting the stored data for an extended period of time. For this reason, Marchese \textit{et al.} proposed a hypothetical energy-aware routing protocol using contact-graph routing (CGR) \cite{Marchese2018}. CGR  accounts for prior contact data, including start time, end time, and overall contact volumes, in order to complete the path from source to destination. In the CGR and extended-CGR methods, CubeSats transmit to the ground station only when they have enough energy for data forwarding. Nevertheless, Challa \textit{et al.} proposed a torrent-based approach  for CubeSats to improve the downlink and uplink times for large file transmissions \cite{Challa2012}. In CubeSat-torrent, large files are split into small chunks,  resulting in  low latency. Liu \textit{et al.} proposed an analytical framework, which formulates the data-cquisition and delivery strategies for small satellites \cite{Liu2016}. They have shown numerically that joint optimization of data-acquisition and data-delivery can improve delay-constrained throughput.
\begin{table*}
\caption{{  Networking for CubeSats}}\label{table:networking}
 \centering
{  \begin{tabular}{|p{1.0cm}|p{3.5cm}|p{2.3cm}|p{7.3cm}| }
\hline
 \hline
    \textbf{Ref.} &\textbf{Protocol}& \textbf{Architecture} &\textbf{Comments}\\
       \hline\hline
       \cite{Radhakrishnan2013}       &Bellman-Ford  &Static& Estimates shortest path for CubeSat constellation \\ \hline
              \cite{BERGAMO2005725}       &  &Static& Routing and synchronization among  CubeSats  \\ \hline
                      \cite{Zhang2011}       & Bandwidth-and-delay-aware routing &Static& Considers both delay and bandwidth aspects for  inter-satellite links  \\ \hline
    \cite{Yang2008}       & Multi-cast routing &Static& Avoids using non-feasible links and minimizing bandwidth  \\ \hline
        \cite{Di2005}      & Dynamic routing &Static& Provides more autonomy, limited overhead, and compatible functionality  \\ \hline
                \cite{Marchese2018}     & Contact-graph routing &Static& Supports delay-tolerant networking  \\ \hline
                                \cite{Challa2012}     & Torrent-based routing &Static& Splits large files into small chunks to reduce latency  \\ \hline
   \cite{Liu2016}     & Joint optimization of data acquisition and delivery &Static& Improves delay-constrained throughput  \\ \hline
      \cite{Sheng2017}     & SDN-based networking& Dynamic &Resource management for SDN-and NFV-based satellite networks \\ \hline
           \cite{FERRUS201695}     & SDN-based networking & Dynamic & SDN- and NFV-based ground segments for satellite networks \\ \hline
                     \cite{AKYILDIZ2019134}     & SDN-based networking & Dynamic & Introduces architecture and protocols for SDN-enabled small-satellite networks \\
                     \hline
                                          \cite{Xu2018}     & SDN-based networking & Dynamic & Fault recovery and mobility management for SDN-enabled small-satellite networks \\ \hline
       \cite{Kak2018}     & SDN-based networking & Dynamic & Design of SDN-based multi-band radio for small-satellite networks \\ \hline
 \hline
 \hline
\end{tabular}}
\end{table*}
The traditional centralized routing, based on static architecture leads to inefficient network performance. To solve this problem, the paradigms of software-defined networking (SDN) and network function virtualization (NFV) are introduced to simplify network management, improve network utilization, and provide fine-grained control for satellite networks. In \cite{Sheng2017}, the authors introduce resource management schemes for SDN and NFV-based  LEO satellite networks with a GEO relay. In \cite{FERRUS201695}, Ferrus \textit{et al.} used the SDN-based architecture for satellite networks but limited to the ground part only. Alternatively, Li \textit{et al.} introduced a multi-level data-forwarding method, in which the ground station handles the  network management while the GEO satellites manage the controlling \cite{Li2018}. This approach is time-consuming with high latency, since it requires  relaying control information from GEO to LEO satellites.

Recently, Akyildiz \textit{et al. }  introduced an end-to-end architecture for SDN/NFV-based small-satellite networks \cite{AKYILDIZ2019134}. Fig.~\ref{fig: sdn-iost} shows this network architecture, which consists of three layers: infrastructure, control-and-management, and policy layers \cite{AKYILDIZ2019134}. The infrastructure layer consists of the CubeSats, on-ground sensing devices, switches, and gateways. The control-and-management layer is mainly responsible for network management, control, and performance optimization. The policy layer, interfaced with the control-and-management layer, is treated as an external layer that provides an abstract view of the network to the user (see Fig.~\ref{fig: sdn-iost}). SDN and NFV simplify  network management, improve network utilization, and provide fine-grained control for the system hardware \cite{Kak2018}.

Xu \textit{et al. } proposed a similar architecture with a fault recovery mechanism and mobility management using SDN \cite{Xu2018}. This network design,  called SoftSpace, is identical to the one in \cite{AKYILDIZ2019134} consisting of four segments, a user segment, a control-and-management segment, a ground segment, and a space segment. The  SoftSpace architecture's data plane consists of a satellite-core network (SCN) and a satellite-access network (SAN) (see Fig.~\ref{fig: sdn-iost}). The SCN consists of SDN switches, whereas SAN comprises satellite gateways, terminals, and satellites. Each SDN-enabled satellite can perform four main functions: i) creating programmable physical and MAC-layer functions; ii) supporting the rules for SDN-based packet handling which can be configured by the network controller through OpenFlow southbound application program interfaces; iii) creating a wireless hypervisor to enable virtual SDN-satellites; and iv) supporting multi-band communication technologies, such as different RF bands and optical bands. Recently, Kak \textit{et al.} propsoed SDN-based multi-band radios,  in which the impact of different carrier frequencies and orbital parameters on the latency and throughput were investigated \cite{Kak2018}. Average end-to-end throughput of $489$ and $35$ Mbps were achieved for mmWaves and the S-band, respectively \cite{Kak2018}. Furthermore, Xu \textit{et al.} addressed the issue of controller placement in SDN-based satellite networking  with a three-layer hierarchical architecture \cite{Xu2018a}. The domain controller, slave controller, and super controllers were placed on the GEO satellites, the LEO satellites, and the ground stations, respectively \cite{Xu2018a}. Table \ref{table:networking} summarizes the literature on CubeSat networking.
}


{  
\section{Application Layer}
Application-layer protocols are vital for providing connectivity to various user applications. The literature on application-layer protocols for small satellites is not very extensive. However, some recent works discuss the use of conventional application-layer protocols in space-information networks to provide connectivity to IoT applications. For instance, in \cite{Bacco2019}, the authors analyze two application-layer protocols, Message Queuing Telemetry Transport (MQTT) and constrained application (CoAP) in space-information networks. The MQTT protocol was initially designed by IBM to support satellite networks; however, it is widely used in terrestrial networks. In the MQTT protocol, the data producer and data consumer are separated by a rendezvous entity, also known as a broker. The data is organized into logical flows (topics), through which it is sent to the broker, which keeps track of both active subscriptions and topics. Since MQTT is a TCP-based protocol, it is reliable and energy-inefficient. Alternatively, CoAP relies on representational state transfer, which supports resource-constrained conditions. The resources are encapsulated in CoAP servers, which are addressable through resource identifiers \cite{Bacco2018}. The CoAP client sends confirmable and non-confirmable requests for a resource query to the server. Unlike MQTT, CoAP uses UDP and therefore has low reliability; however, it is more energy-efficient with low overhead. Due to TCP bandwidth probing, MQTT has larger variations in good-put comparing to the CoAP protocol  \cite{Bacco2019}.
}


\section{Future Research Directions}
We envision the use of CubeSats to enable various applications of future wireless communications in space. Compared to the current satellite communication systems, CubeSats have many attractive features, such as low cost and low orbital altitudes. However, the research on CubeSats for communications is still in its early phase, and therefore poses a wide variety of problems. In this section, we point out the significant research challenges facing CubeSat communications.

\subsection{Integration with Next Generation Wireless  Systems}
One interesting research area is the integration of CubeSats communications with next-generation wireless networks, including 5G and beyond \cite{TarKhaBennis:19, sarieddeen2019next}. For example, Babich \textit{et al.} have recently introduced an integrated architecture for nano-satellites and 5G \cite{Babich2019}. The terrestrial communication link operates on mmWaves, while the satellite-to-ground links and inter-satellite links use RF communication. For 5G and beyond systems, the intrinsic ubiquity and long-coverage capabilities of CubeSats would make them major candidates for overcoming the digital divide problem in future wireless communications. As  existing studies are limited in this  area, investigating ways of integrating CubeSats with next-generation networks, at both the physical and networking layers, is a promising research avenue for achieving the ambitious data metrics of 5G and beyond systems.
\subsection{Scheduling}
Because of their small size, CubeSats have a limited number of onboard transceivers, which in turn limits the number of communication contacts. Therefore, data scheduling is necessary to employ the available transceivers efficiently. Hence, in \cite{Zhou2019}, Zhou \textit{et al.} proposed a finite-embedded-infinite two-level programming technique to schedule the data for CubeSats optimally. This technique considers stochastic data arrival and takes into account the joint considerations of battery management, buffer management, and contact selection. This framework demonstrats a significant gain in the downloaded data on battery and storage capacities. Similarly, Nag \textit{et al.} designed a scheduling algorithm  for CubeSats, that had four-times better computational speed than  integer programming in \cite{NAG2018891}. The optimized scheduler incorporates the attitude control and orbital mechanics of the CubeSats to maximize their coverage. These theoretical scheduling models need to be mission-tested for validation. Also, these scheduling frameworks can be integrated with the UAVs and high-altitude platforms (HAPs) to create a more complete space-information network.

From another perspective, a scheduling method for  necessary CubeSat tasks is proposed in  \cite{SLONGO2018141}. This scheduling algorithm opts for the number and the type of tasks to be executed in order to permit the energy harvesting  system's solar panels  to operate close to their maximum power point, leading to higher power efficiency \cite{SLONGO2018141}.  This can lead to around a 5\% decrease in energy consumption compared to systems without a task scheduler \cite{SLONGO2018141}. As a future research direction, we propose investigating the problem of joint scheduling and data routing among multiple CubeSats, especially in cases that favor particular CubeSats-to-ground communication routes.

\subsection{Software-Defined Networking}
The existing architectures of broadband satellite communication networks are inflexible due to their dependence on hardware. However, some recent works (\cite{Rossi2015}, \cite{FERRUS201695}, and \cite{Rossi2018}) introduce the concept of using SDN for broadband satellite communication networks to improve their flexibility. In addition to its use in broadband satellite communications, SDN has also been recently proposed for CubeSat communications. For example, Kak \textit{et al.} used SDN and network function virtualization (NFV) to provide connectivity to IoT networks \cite{Kak2018}. Akyildiz \textit{et al.} showed  that SDN and NFV improve network utilization and hardware control, and simplify  network management \cite{AKYILDIZ2019134}. However, Vital \textit{et al.} showed that implementing SDN/NFV for CubeSat communications poses multiple technical challenges \cite{Vital2020}. Questions such as how SDN protocols can be applied to  CubeSat gateways and remote terminals, how to perform dynamic network configurations to meet QoS demand, and how to provide on-demand services without affecting regular network operation, remain open at the moment, and offer promising avenues for future research.

\subsection{Towards Internet of Space Things}
NASA aims to establish a human colony on Mars by 2025, which will require connectivity beyond Earth \cite{skyworld}. To provide such intra-galactic connectivity, the Internet of space things (IoST) is an enabling technology consisting of deep-space CubeSats. To that end, there is a growing interest in the space industry to establish IoST networks, which are still in the early development phase. Besides, IoST will also provide extended coverage to on-ground cyber-physical systems in rural areas. For instance,  with its $66$ small satellites, also called SensorPODs, Iridium Communications offers connectivity solutions for Earth-remote sensing and space research \cite{gupta2011}. CubeSats will play a significant role in the development of IoST networks, where inter-satellite communication, in-space backhauling, and data forwarding are some of the exciting challenges.

\subsection{Hybrid Architecture}
CubeSats can also be integrated with other communication technologies, including GEO and MEO satellites, HAPs, and UAVs. For example, in space-information networks, CubeSats act as relays between the GEO and MEO satellites. Similarly, CubeSats can perform back-hauling for HAPs and high-altitude UAVs. Some recent works, such as \cite{Du2016}, discusses a hybrid architecture for CubeSats, conventional satellites, HAPs, and UAVs. However, these architectures are purely hypothetical models requiring further validation. Also, inter-linking  these entities is challenging because of the dynamic nature of these technologies and the uncertain space environment.

\subsection{LoRa for CubeSats}
The typical \ac{IoT} scenario involves connecting devices with limited energy over long ranges. In this regard, terrestrial-based \acp{LPWAN} aim to offer low-data-rate communication capabilities over a wide area. An important communication technique for terrestrial \ac{IoT} networks, the \ac{LoRaWAN}, is based on a novel frequency-shift chirp-spread-spectrum-modulation technique called LoRa\cite{SelOliSor:16,alliance2015technical,ChiElza:19}.  Unfortunately, terrestrial-based \acp{LPWAN}, including \ac{LoRaWAN}, cannot offer ubiquitous coverage, especially for remote areas, e.g., desert, forests, and farms, mainly  for economic reasons. On the contrary, IoT LEO nanosatellites can serve as a cost-efficient solution to this problem by providing global coverage. However, the modulation and multiple-access techniques usually adopted in terrestrial \ac{IoT} systems {  cannot be directly} used in  CubeSats, due to the Doppler effect and  propagation-delay constraints.

In this regard, an architecture for an \ac{IoT}-based satellite system consisting of a LEO Flower constellation made up of $35$ satellites in seven orbital planes with two additional polar satellites, was outlined in \cite{Qu2017} to assure global coverage. The compatibility of the communication protocols between the terrestrial \ac{IoT} systems, e.g., LoRa and narrowband-\ac{IoT}, and their satellite counterparts were also discussed.   It was shown that either cognitive radio mechanisms for interference mitigation or spread-spectrum techniques should be used to permit the coexistence of both terrestrial and satellite networks with an acceptable interference level. It was also shown that modifications of the existing higher-layer protocols for \ac{IoT} systems are required to decrease the overhead data to cope with limited power and  delay involved in  satellite-based networks.

In \cite{Doroshkin2019}, the authors have shown the feasibility of LoRa modulation in CubeSats, where the Doppler effect may have a non-negligible impact on the performance.
It was found experimentally that at higher orbits with altitudes more than $550$ km, LoRa modulation is immune to the Doppler effect. By contrast,  rapid variations in the Doppler frequency shift for instance, when a lower altitude satellite flies directly above the ground station, lead to severe degradation in the performance, which reduces the duration of the radio communication session. From another perspective, after some modification, the frequency shift spread-spectrum in LoRa  can be used to provide a multiple-access technique as an alternative to the direct-sequence spread spectrum traditionally used in satellites \cite{QiaMaLia:19}. It was shown that the \ac{BER} performance of the proposed frequency-and-phase-symmetry chirp-spread spectrum are similar to the direct-sequence multiple access.

\subsection{Machine Learning for Resource Allocation in CubeSats}
It is fair to believe that more intelligence will be added to future wireless communication networks using deep-learning techniques. A pressing challenge faced by CubeSat communications, is the limited bandwidth that lead to low data rates, high latency, and eventual performance degradation. Therefore, we envision equipping CubeSats with multi-band connectivity and smart capabilities to allocate power and spectrum resources in a dynamic manner across microwave, mmWaves, THz-band, and optical frequencies \cite{Nie2019}.
This adaptive solution requires new transceivers and antenna systems, which are challenging research directions. Moreover, investigating and developing the performance of new resource-allocation schemes will be the result of customized machine-learning strategies.  For example, Nie \textit{et al.} proposed a new multi-objective resource-allocation scheme based on deep neural network (DNN) \cite{Nie2019}. Instead of the back-propagation algorithm, the random hill-climbing algorithm was utilized to adjust the weights of the neurons. This study was based on real satellite trajectory data from Iridium NEXT small satellites examining the influences of the Doppler shift and heavy-rain fade. The proposed DNN-based scheme resulted in improved multi-Gbps throughput for the inter-satellite links, and can be adopted in a multitude of future CubeSat communication problems with similar structures.

\section{Conclusions}
We envision CubeSats enabling a wide range of applications, including Earth and space exploration, rural connectivity for the pervasive Internet of things (IoT) networks, and ubiquitous coverage.
Current CubeSat research is mostly focused on remote-sensing applications. Unfortunately, few efforts have been made to offer communication solutions using CubeSats, which could involve CubeSat swarms for ubiquitous coverage, optical communication for high data rates, integration with future cellular systems for back-hauling, etc. Therefore, in this paper, we have reviewed the literature on various facets of CubeSat communications, including channel modeling, modulation and coding, coverage, networking, and upper-layer issues. We also outlined several significant future research challenges,  that highlight how CubeSat technology is a key enabler for the emerging Internet of space things.
Both the existing literature collection and the research problems we propose  form a promising framework for addressing the global problem of the digital divide. In short, this paper is a good starting point for the academic and industrial researchers focusing on providing communication solutions using CubeSats.

%
%
%
%
\bibliographystyle{../bib/IEEEtran}
\bibliography{../bib/IEEEabrv,../bib/nasir_ref}

\vspace{-3.5em}
\begin{IEEEbiography}[{\includegraphics[width=1in,height=1.25in]{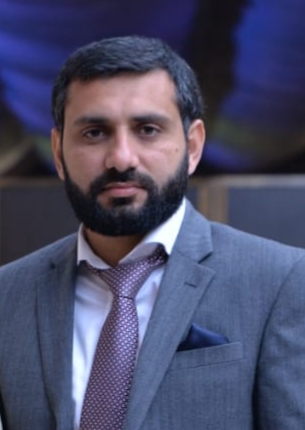}}]{Nasir Saeed}(S'14-M'16) received his Bachelors of Telecommunication degree from University of Engineering and Technology, Peshawar, Pakistan, in 2009 and received Masters degree in satellite navigation from Polito di Torino, Italy, in 2012. He received his Ph.D. degree in electronics and communication engineering from Hanyang University, Seoul, South Korea in 2015. He was an Assistant Professor at the Department of Electrical Engineering, Gandhara Institute of Science and IT, Peshawar, Pakistan from August 2015 to September 2016. Dr. Saeed worked as an assistant professor at IQRA National University, Peshawar, Pakistan from October 2017 to July 2017. He is currently a Postdoctoral Research Fellow in King Abdullah University of Science and Technology (KAUST).   His current areas of interest include cognitive radio networks, underwater and underground  wireless communications, satellite communications, dimensionality reduction, and localization.
\end{IEEEbiography}
\vspace{-2.5em}
\begin{IEEEbiography}[{\includegraphics[width=1in,height=1.25in]{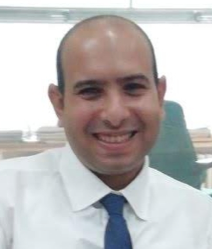}}]{Ahmed Elzanaty}(S’13-M’19) received the Ph.D. degree (excellent cum laude) in Electronics, Telecommunications, and Information technology from the University of Bologna, Italy, in 2018.  He was a research fellow at the University of Bologna from 2017 to 2019. Currently, he is a post-doctoral fellow at King Abdullah University of Science and Technology (KAUST), Saudi Arabia. He has participated in several national and European projects, such as GRETA and EuroCPS. His research interests include statistical signal processing, digital communications, and information theory. He is the recipient of the best paper award at the IEEE Int. Conf. on Ubiquitous Wireless Broadband (ICUWB 2017). Dr. Elzanaty was a member of the Technical Program Committee of the European Signal Processing Conf. (EUSIPCO 2017 and 2018). He is also a representative of the IEEE Communications Society’s Radio Communications Technical Committee for several international conferences.
\end{IEEEbiography}

\begin{IEEEbiographynophoto}{Heba Almorad}(S’19) is an  electrical and computer engineering undergraduate student at Effat University, Saudi Arabia. She is expected to graduate and receive her B.S. degree  in 2021. Her current areas of interest include satellite communications, Cubesats, antenna design, and machine learning.
\end{IEEEbiographynophoto}

\begin{IEEEbiography}[{\includegraphics[width=1in,height=1.25in]{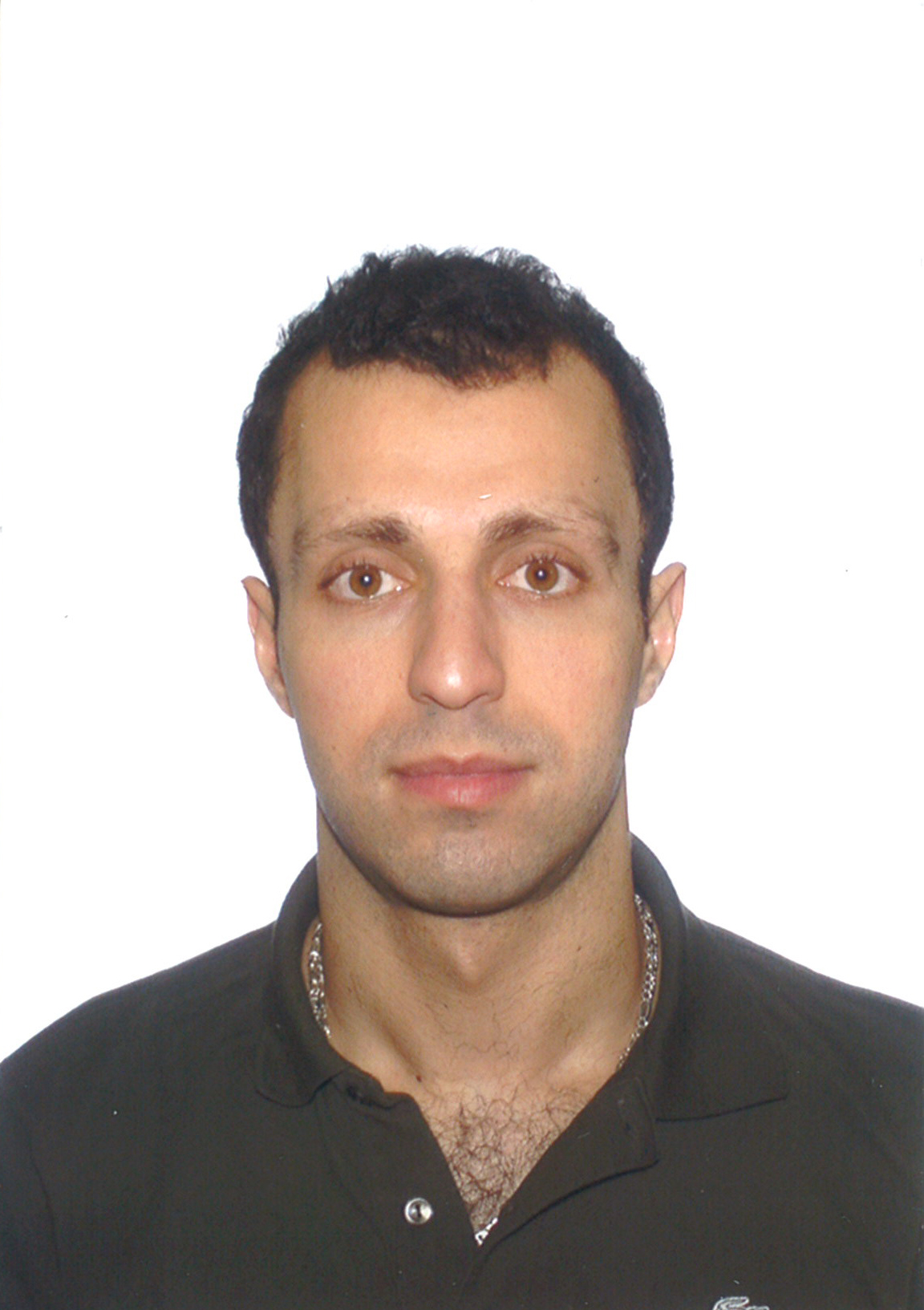}}]{Hayssam Dahrouj}(S'02, M'11, SM'15) received his B.E. degree (with high distinction) in computer and communications engineering from the American University of Beirut (AUB), Lebanon, in 2005, and his Ph.D. degree in electrical and computer engineering from the University of Toronto (UofT), Canada, in 2010. In May 2015, he joined the Department of Electrical and Computer Engineering at Effat University as an assistant professor, and also became a visiting scholar at King Abdullah University of Science and Technology (KAUST). Between April 2014 and May 2015, he was with the Computer, Electrical and Mathematical Sciences and Engineering group at KAUST as a research associate. Prior to joining KAUST, he was an industrial postdoctoral fellow at UofT, in collaboration with BLiNQ Networks Inc., Kanata, Canada, where he worked on developing practical solutions for the design of non-line-of sight wireless backhaul networks. His contributions to the field led to five patents. During his doctoral studies at UofT, he pioneered the idea of coordinated beamforming as a means of minimizing intercell interference across multiple base stations. The journal paper on this subject was ranked second in the 2013 IEEE Marconi paper awards in wireless communications. His main research interests include cloud radio access networks, cross-layer optimization, cooperative networks, convex optimization, distributed algorithms, satellite communications, and free-space optical communications.
\end{IEEEbiography}
\begin{IEEEbiography}[{\includegraphics[width=1in,height=1.25in]{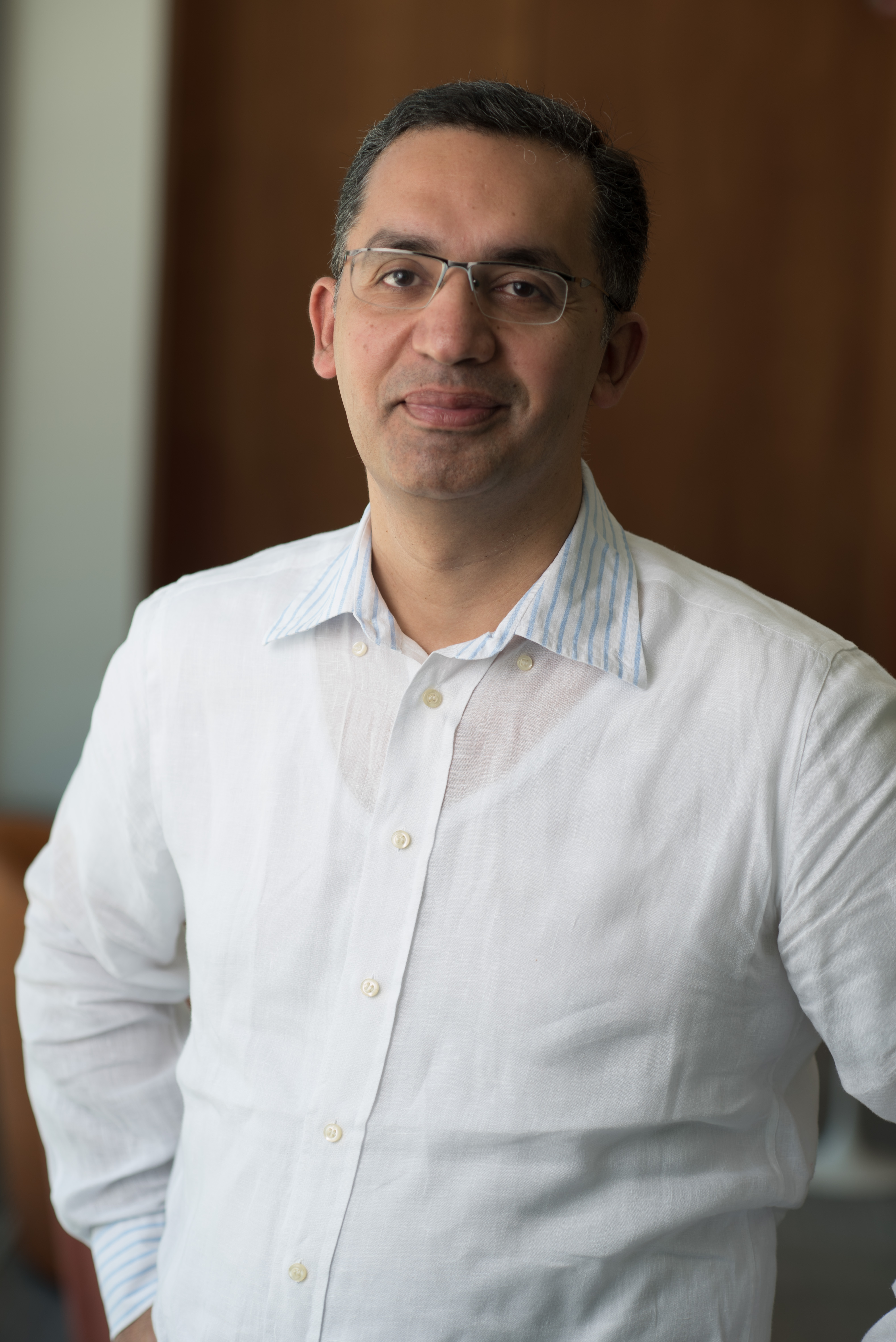}}]{Tareq Y. Al-Naffouri }
(M'10-SM'18) Tareq  Al-Naffouri  received  the  B.S.  degrees  in  mathematics  and  electrical  engineering  (with  first  honors)  from  King  Fahd  University  of  Petroleum  and  Minerals,  Dhahran,  Saudi  Arabia,  the  M.S.  degree  in  electrical  engineering  from  the  Georgia  Institute  of  Technology,  Atlanta,  in  1998,  and  the  Ph.D.  degree  in  electrical  engineering  from  Stanford  University,  Stanford,  CA,  in  2004.
He  was  a  visiting  scholar  at  California  Institute  of  Technology,  Pasadena,  CA  in  2005  and  summer  2006.  He  was  a  Fulbright scholar  at  the  University  of  Southern  California  in  2008.  He  has  held  internship  positions  at  NEC  Research  Labs,  Tokyo,  Japan,  in  1998,  Adaptive  Systems  Lab,  University  of  California  at  Los  Angeles  in  1999,  National  Semiconductor,  Santa  Clara,  CA,  in  2001  and  2002,  and  Beceem  Communications  Santa  Clara,  CA,  in  2004.  He  is  currently  a  Professor  at  the  Electrical  Engineering  Department,  King  Abdullah  University  of  Science  and  Technology  (KAUST).  His  research  interests  lie  in  the  areas  of  sparse, adaptive,  and  statistical  signal  processing  and  their  applications,  localization,  machine  learning,  and  network  information  theory.    He  has  over  240  publications  in  journal  and  conference  proceedings,  9  standard  contributions,  14  issued  patents,  and  8  pending.  Dr.  Al-Naffouri  is  the  recipient  of  the  IEEE  Education  Society  Chapter  Achievement  Award  in  2008  and  Al-Marai  Award  for  innovative  research  in  communication  in  2009.  Dr.  Al-Naffouri  has  also  been  serving  as  an  Associate  Editor  of  Transactions  on  Signal  Processing  since  August  2013.
\end{IEEEbiography}
%
\begin{IEEEbiography}[{\includegraphics[width=1in,height=1.25in]{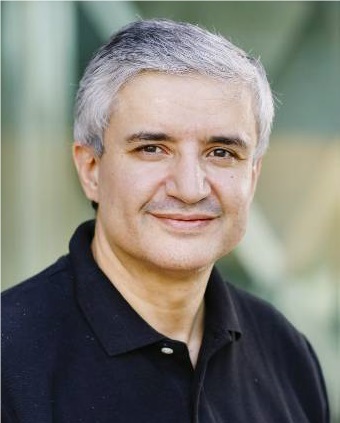}}]{Mohamed-Slim Alouini}
(S'94-M'98-SM'03-F'09)  was born in Tunis, Tunisia. He received the Ph.D. degree in Electrical Engineering
from the California Institute of Technology (Caltech), Pasadena,
CA, USA, in 1998. He served as a faculty member in the University of Minnesota,
Minneapolis, MN, USA, then in the Texas A\&M University at Qatar,
Education City, Doha, Qatar before joining King Abdullah University of
Science and Technology (KAUST), Thuwal, Makkah Province, Saudi
Arabia as a Professor of Electrical Engineering in 2009. His current
research interests include the modeling, design, and
performance analysis of wireless communication systems.
\end{IEEEbiography}

\end{document}